\documentclass{aastex62}
\usepackage{amsmath}

\graphicspath{{figures/}}

\shorttitle{PREPROCESSING AMOUNG EDISCS CLUSTER GALAXIES}
\shortauthors{JUST ET AL.}
\newcommand{\tx}[1]{\textrm{#1}}

\newcommand{\kms}{km~$\tx{s}^{-1}$}
\newcommand{\msun}{$\ensuremath{M_{\odot}}$}

\def\gtsim{\lower 2pt \hbox{$\, \buildrel {\scriptstyle >}\over
    {\scriptstyle \sim}\,$}}
\def\ltsim{\lower 2pt \hbox{$\, \buildrel {\scriptstyle <}\over
    {\scriptstyle \sim}\,$}}

\begin{document}
\title{Preprocessing Among the Infalling Galaxy Population of EDisCS Clusters}

%%% BEGIN AUTHORS %%%
\author{Dennis~W.~Just}
\affiliation{Steward Observatory, University of Arizona, 933 North Cherry Avenue, Tucson, AZ 85721, USA}
\affiliation{Department of Astronomy \& Astrophysics, University of Toronto, 50 St George Street, Toronto, ON M5S 3H4, Canada}

\author{Matthew~Kirby}
\affiliation{Department of Physics, University of Arizona, 1118 E 4th Street, Tucson, AZ 85721, USA}

\author{Dennis~Zaritsky}
\affiliation{Steward Observatory, University of Arizona, 933 North Cherry Avenue, Tucson, AZ 85721, USA}

\author{Gregory~Rudnick}
\affiliation{Deptartment of Physics and Astronomy, University of Kansas, Lawrence, KS 66045, USA}

\author{Tyler Desjardins}
\affiliation{Deptartment of Physics and Astronomy, University of Kansas, Lawrence, KS 66045, USA}

\author{Richard~Cool}
\affiliation{MMT Observatory, 1540 E. Second Street, University of Arizona, Tucson, AZ 85721, USA}

\author{John~Moustakas}
\affiliation{Department of Physics \& Astronomy, Siena College, 515 Loudon Road, Loudonville, NY 12211, USA}

\author{Douglas~Clowe}
\affiliation{Department of Physics \& Astronomy, Ohio University, Clippinger Labs 251B, Athens, OH 45701, USA}

\author{Gabriella~De~Lucia}
\affiliation{INAF - Osservatorio Astronomico di Trieste, via Tiepolo 11, I-34143, Trieste, Italy}

\author{Alfonso~Arag{\'o}n-Salamanca}
\affiliation{School of Physics and Astronomy, The University of Nottingham, University Park, Nottingham NG7 2RD, UK}

\author{Vandana~Desai}
\affiliation{IPAC, Mail Code 100-22, Caltech, 1200 E. California Blvd. Pasadena, CA 91125, USA}

\author{Rose~Finn}
\affiliation{Department of Physics, Siena College, Loudonville, NY, USA}

\author{Claire~Halliday}
\affiliation{23, rue d'Yerres, F-91230 Montgeron, France}

\author{Pascale~Jablonka}
\affiliation{GEPI, Observatoire de Paris, PSL University, CNRS, 5 Place Jules Janssen, F-92190 Meudon, France}
\affiliation{Institute of Physics, Laboratoire d’Astrophysique, Ecole Polytechnique F\'ed\'erale de Lausanne (EPFL), Observatoire, 1290 Versoix, Switzerland}

\author{Justin Mann}
\affiliation{Deptartment of Physics and Astronomy, University of Kansas, Lawrence, KS 66045, USA}

\author{Bianca~Poggianti}
\affiliation{INAF - Astronomical Observatory of Padova, vicolo dell'Osservatorio 5, Padova I-35122, Italy}

\author{Fu-Yan~Bian}
\affiliation{European Southern Observatory, Alonso de C\'ordova 3107, Casilla 19001, Vitacura, Santiago 19, Chile}

\author{Kelley~Liebst}
\affiliation{School of Earth and Space Exploration, Arizona State University, 781 E Terrace Mall, Tempe, AZ 85287, USA}
%%% END AUTHORS %%%

\begin{abstract}
We present results from a low-resolution spectroscopic survey for 21
galaxy clusters at $0.4<z<0.8$ selected from the ESO Distant Cluster
Survey. We measured spectra using the low-dispersion prism in
IMACS on the Magellan Baade telescope and calculate redshifts with an accuracy of
$\sigma_z=0.007$. We find 1763 galaxies that are brighter than $R=22.9$ in
the large-scale cluster environs. We identify the galaxies expected to be
accreted by the clusters as they evolve to \hbox{$z=0$} using
spherical infall models and find that $\sim30$--$70\%$ of the $z=0$ cluster population
lies outside the virial radius at $z\sim0.6$. For analogous clusters at $z=0$, we
calculate that the ratio of galaxies that have fallen into the clusters since
$z\sim0.6$ to those that were already in the core at that redshift is typically
between $\sim0.3$ and $1.5$. This wide range of ratios is due to intrinsic
scatter and is not a function of velocity dispersion, so a variety of infall histories
is to be expected for clusters with current velocity dispersions of
$300\ltsim\sigma\ltsim 1200$~km~s$^{-1}$. Within the infall regions of $z\sim0.6$
clusters, we find a larger red fraction of galaxies than in the
field and greater clustering among red galaxies than blue. We
interpret these findings as evidence of ``preprocessing,'' where galaxies in denser
local environments have their star formation rates affected prior to their aggregation
into massive clusters, although the possibility of backsplash galaxies complicates
the interpretation.
\end{abstract}

%%%%%%%%%%%%%%%%%%%%%%%%%%%%%%%%%%%
\section{Introduction}\label{sec:intro}

Although a relationship exists between the evolution of galaxies and
their environment, as demonstrated by correlations between density and
galaxy color \citep[e.g.,][]{Hogg04}, star-formation \citep{Lewis02,Gomez03},
and morphology \citep{Dressler80,Postman84}, the physical processes that drive
these changes and the connection between those processes and
environment are not established. While the cores of clusters are the
final resting place for quiescent galaxies and are where these trends were
discovered, the key to understanding the implicit quenching
of star formation and morphological transformation is to study
galaxies in the environment where they are being transformed, not
where they ultimately reside.

Quenching and morphological transformation do not occur primarily in
the cores of clusters, at least not at redshifts $<1$. The decrease in star
formation sets in at several virial radii \citep{Lewis02,Gomez03}, and the increase
in the S0 fraction since $z\sim0.5$ \citep{Dressler97,Fasano00,Desai07} is most dramatic
in less massive clusters \citep{Poggianti09,Just10}. Environmentally driven evolution
occurs primarily at intermediate densities, which should include the environs outside
the cluster virial radius. Such effects are predicted in simulations out to as many as
five virial radii \citep{Bahe13}. Establishing the size and characteristics of the infalling
galaxy population will therefore constrain the path to transformation.

As a result of this line of thought, a number of studies have begun to
target the outskirts of $z\gtsim0.5$ massive clusters
\citep[e.g.,][]{Moran07,Patel11,Oemler13}. However, such studies have been limited to
a few clusters ($\sim 10$), making general conclusions difficult to reach
given the variation in properties from cluster to cluster.  Because of
the high masses of these targeted clusters and correspondingly large
virial radii, some of these studies do not probe very far past the
virial radius and may miss a significant fraction of the infalling galaxies.
Furthermore, such clusters are also rare; hence, the infalling population of more
typical clusters has not been explored. This bias may lead to an incomplete picture,
given the cluster mass dependence of S0 evolution \citep{Poggianti09,Just10}.

A fundamental difficulty in studying cluster infalling populations is
the contamination of interloping foreground/background galaxies, an issue that
becomes more important at larger clustercentric radii where the relative
fraction of interlopers is larger. The studies listed above use
spectroscopic redshifts for this purpose, but this approach requires
significant telescope time and is thus limited to those few clusters.
The alternative approach using photometric redshifts \citep[e.g.,][]{Kodama01} comes
with much lower observational cost, but photometric redshifts are insufficiently precise
to securely associate a galaxy with a particular cluster, where
$c\delta z\approx 500$~km~s$^{-1}$ resolution is needed.

We adopt a hybrid approach. We isolate the infalling galaxy population of $21$ clusters
at $0.4<z<0.8$ using the Low-Dispersion Prism
(LDP\footnote{Designed by S. Burles for use by the PRIMUS redshift survey
\citep{Coil11}.}) installed in the Inamori-Magellan Areal Camera and Spectrograph
\citep[IMACS;][]{Bigelow98,Dressler06} on the 6.5m Magellan Baade telescope.
With these data, we measure the number of galaxies these clusters will
accrete by $z=0$ to establish how many galaxies may be influenced by the
accretion process. We also measure the scatter in this number to estimate
the range in accretion histories. We compare models \citep[e.g.,][]{Poggianti06}
that predict the amount of mass accreted by these clusters to our
observations. Finally, we measure the optical properties and clustering
amplitude of infalling galaxies to quantify the amount of evolution that takes
place outside the virial radius.

This paper is organized as follows. In \S2 we describe our sample selection
and the sample's basic properties, and in \S3 and \S4 we present the imaging
and spectroscopic data, respectively. In \S5 we analyze our clusters using
mass infall models and quantify the number of galaxies and optical properties
of the infalling population. We conclude in \S6. All magnitudes in this paper are in
the AB system; to convert these to the Vega system, subtract $0.02$, $0.06$,
$0.23$, $0.45$, and $0.55$ from the AB magnitudes for the $BVRIz$ bands,
respectively. Throughout the paper, we adopt $H_{0}=70$~\kms~Mpc$^{-1}$,
$\Omega_{0}=0.3$, and $\Omega_{\Lambda}=0.7$, and all cosmology-dependent
quantities taken from other studies also use these values. We approximate the
virial radii of our clusters as $R_{\rm 200}$, the radius inside which the
enclosed density is $200$ times the critical density of the universe at that redshift.

%%%%%%%%%%%%%%%%%%%%%%%%%%%%%%%%%%%
\section{Sample}\label{sec:sample}

Our sample consists of $21$ galaxy clusters. We include $16$ of the $20$ galaxy
clusters in the ESO Distant Cluster Survey \citep[EDisCS;][]{White05}; see
\S\ref{sec:ldptargets} for details of the four clusters not observed. We also
include the seven clusters found serendipitously in this survey. Of these $23$
LDP-observed clusters, two are removed from the analysis for reasons given in
\S\ref{sec:ldpz}, resulting in $21$ clusters in our final sample. We present
basic information on the clusters in Table~1.

%%
%% TABLE 1
%%
\begin{deluxetable}{rlcccccclc}
\tablecolumns{10}
\tablenum{1}
\tabletypesize{\scriptsize}
\tablewidth{0pt}
\tablecaption{LDP-Observed EDisCS Clusters}
\tablehead{
\colhead{Field} &
\colhead{Cluster ID} &
\colhead{R.A.} &
\colhead{Decl.} &
\colhead{$z$} &
\colhead{$\sigma$} &
\colhead{$R_{200}$} &
\colhead{$M_{\rm 200}$} &
\colhead{Imaging} &
\colhead{Seeing ($\arcsec$)} \\
\colhead{(1)} &
\colhead{(2)} &
\colhead{(3)} &
\colhead{(4)} &
\colhead{(5)} &
\colhead{(6)} &
\colhead{(7)} &
\colhead{(8)} &
\colhead{(9)} &
\colhead{(10)}
}
\startdata
1  & Cl1018.8$-$1211   & 10:18:46.8 & $-$12:11:53 & 0.4734 &  $486^{+59}_{-63}$   & $0.93^{+0.11}_{-0.12}$ &  $1.53^{+0.63}_{-0.52}$(14) & $VRI$   & $1.20$ \\
2  & Cl1037.9$-$1243   & 10:37:51.2 & $-$12:43:27 & 0.5783 &  $319^{+53}_{-52}$   & $0.58^{+0.10}_{-0.09}$ &  $4.06^{+2.38}_{-1.68}$(13) & $BVRIz$ & $2.10$ \\
3  & Cl1037.9$-$1243a  & 10:37:52.3 & $-$12:44:49 & 0.4252 &  $537^{+46}_{-48}$   & $1.06^{+0.09}_{-0.09}$ &  $2.12^{+0.59}_{-0.52}$(14) & $BVRIz$ & $2.10$ \\
4  & Cl1040.7$-$1155   & 10:40:40.4 & $-$11:56:04 & 0.7043 &  $418^{+55}_{-46}$   & $0.70^{+0.09}_{-0.08}$ &  $8.47^{+3.80}_{-2.50}$(13) & $BVRIz$ & $1.45$ \\
5  & Cl1054.4$-$1146   & 10:54:24.5 & $-$11:46:20 & 0.6972 &  $589^{+78}_{-70}$   & $0.99^{+0.13}_{-0.12}$ &  $2.38^{+1.08}_{-0.75}$(14) & $BVRIz$ & $1.20$ \\
6  & Cl1054.7$-$1245   & 10:54:43.6 & $-$12:45:52 & 0.7498 &  $504^{+113}_{-65}$  & $0.82^{+0.18}_{-0.11}$ &  $1.44^{+1.21}_{-0.49}$(14) & $BVRIz$ & $1.25$ \\
7  & Cl1059.2$-$1253   & 10:59:07.1 & $-$12:53:15 & 0.4564 &  $510^{+52}_{-56}$   & $0.99^{+0.10}_{-0.11}$ &  $1.78^{+0.60}_{-0.53}$(14) & $VRI$   & $1.05$ \\
8  & Cl1103.7$-$1245a  & 11:03:34.9 & $-$12:46:46 & 0.6261 &  $336^{+36}_{-40}$   & $0.59^{+0.06}_{-0.07}$ &  $4.61^{+1.65}_{-1.46}$(13) & $BVRI$  & $1.15$ \\
9  & Cl1103.7$-$1245b  & 11:03:36.5 & $-$12:44:22 & 0.7031 &  $252^{+65}_{-85}$   & $0.42^{+0.11}_{-0.14}$ &  $1.86^{+1.84}_{-1.32}$(13) & $BVRI$  & $1.15$ \\
10 & Cl1138.2$-$1133   & 11:38:10.3 & $-$11:33:38 & 0.4796 &  $732^{+72}_{-76}$   & $1.40^{+0.14}_{-0.15}$ &  $5.20^{+1.69}_{-1.46}$(14) & $BVRI$  & $1.15$ \\
11 & Cl1216.8$-$1201   & 12:16:45.1 & $-$12:01:18 & 0.7943 & $1018^{+73}_{-77}$   & $1.61^{+0.12}_{-0.12}$ &  $1.16^{+0.27}_{-0.24}$(15) & $BVRI$  & $1.20$ \\
12 & Cl1227.9$-$1138   & 12:27:58.9 & $-$11:35:13 & 0.6357 &  $574^{+72}_{-75}$   & $1.00^{+0.13}_{-0.13}$ &  $2.29^{+0.97}_{-0.78}$(14) & $BVRI$  & $1.25$ \\
13 & Cl1227.9$-$1138a  & 12:27:52.1 & $-$11:39:59 & 0.5826 &  $341^{+42}_{-46}$   & $0.61^{+0.08}_{-0.08}$ &  $4.95^{+2.06}_{-1.74}$(13) & $BVRI$  & $1.25$ \\
14 & Cl1232.5$-$1250   & 12:32:30.5 & $-$12:50:36 & 0.5414 & $1080^{+119}_{89}$   & $1.99^{+0.22}_{-0.16}$ &  $1.61^{+0.59}_{-0.37}$(15) & $VRIz$  & $1.05$ \\
15 & Cl1301.7$-$1139   & 13:01:40.1 & $-$11:39:23 & 0.4828 &  $687^{+82}_{-86}$   & $1.31^{+0.16}_{-0.16}$ &  $4.29^{+1.73}_{-1.42}$(14) & $VRI$   & $1.15$ \\
16 & Cl1301.7$-$1139a  & 13:01:35.1 & $-$11:38:36 & 0.3969 &  $391^{+63}_{-69}$   & $0.78^{+0.13}_{-0.14}$ &  $8.32^{+4.70}_{-3.67}$(13) & $VRI$   & $1.15$ \\
17 & Cl1353.0$-$1137   & 13:53:01.7 & $-$11:37:28 & 0.5882 &  $666^{+136}_{-139}$ & $1.19^{+0.24}_{-0.25}$ &  $3.67^{+2.74}_{-1.85}$(14) & $VRI$   & $1.20$ \\
18 & Cl1354.2$-$1230   & 13:54:09.7 & $-$12:31:01 & 0.7620 &  $648^{+105}_{-110}$ & $1.05^{+0.17}_{-0.18}$ &  $3.05^{+1.74}_{-1.30}$(14) & $BVRIz$ & $1.66$ \\
19 & Cl1354.2$-$1230a  & 13:54:11.4 & $-$12:30:45 & 0.5952 &  $433^{+95}_{-104}$  & $0.77^{+0.17}_{-0.19}$ &  $1.00^{+0.82}_{-0.56}$(14) & $BVRIz$ & $1.66$ \\
20 & Cl1411.1$-$1148   & 14:11:04.6 & $-$11:48:29 & 0.5195 &  $710^{+125}_{-133}$ & $1.33^{+0.23}_{-0.25}$ &  $4.63^{+2.90}_{-2.15}$(14) & $VRI$   & $1.45$ \\
21 & Cl1420.3$-$1236   & 14:20:20.0 & $-$12:36:30 & 0.4962 &  $218^{+43}_{-50}$   & $0.41^{+0.08}_{-0.09}$ &  $1.36^{+0.97}_{-0.74}$(13) & $VRI$   & $1.00$ \\ \\
\hline \\
22 & Cl1103.7$-$1245   & 11:03:43.4 & $-$12:45:34 & 0.9586 &  $534^{+101}_{-120}$ & $0.77^{+0.15}_{-0.17}$ &  $1.52^{+1.04}_{-0.81}$(14) & $BVRI$  & $1.15$ \\
23 & Cl1138.2$-$1133a  & 11:38:08.6 & $-$11:36:55 & 0.4548 &  $542^{+63}_{-71}$   & $1.05^{+0.12}_{-0.14}$ &  $2.14^{+0.84}_{-0.74}$(14) & $BVRI$  & $1.15$ \\
\enddata
\tablecomments{(1) cluster field; (2) cluster name; (3,4) J2000 R.A. (hr) and decl. (deg); (5) cluster
  redshift \citep{Halliday04,MilvangJensen08}; (6) cluster velocity dispersion in
  km~s$^{-1}$ \citep{Halliday04,MilvangJensen08}; (7) cluster virial
  radius in Mpc; (8) cluster virial mass in units of \msun\ with power of 10 in
  parentheses \citep[using Equation~10 of][]{Finn05};
  (9) wide-field imaging bands observed in each field; (10) the effective WFI seeing after
  smoothing the images to match the band with the poorest seeing for that cluster}
\end{deluxetable}

The EDisCS clusters were drawn from candidates in the Las Campanas Distant
Cluster Survey \citep{Gonzalez01} identified as surface brightness
enhancements in the image background. They lie in a band from $\approx
10$--$14$~hr in R.A. and $\approx -13$ to $-11$~degrees in
decl. They span a redshift range from $z=0.4$ to $0.8$ and
cover a spread in velocity dispersion ($\sigma$) in the range of
$\approx 200$--$1200$~km~s$^{-1}$ \citep{Halliday04,MilvangJensen08},
a wider range of $\sigma$ than other cluster samples at these redshifts
and more representative of the progenitors of $z\sim0$ clusters
\citep{MilvangJensen08}.

We have a variety of data on the cluster cores (the central $\approx6.5'
\times 6.5'$ field of view [FOV]), with deep ($I\ltsim 25$) optical imaging from the FOcal
Reducer and low dispersion Spectrograph (FORS2) on the Very Large Telescope \citep[VLT;][]{White05}, near-infrared (NIR) imaging from the Son OF ISAAC (SOFI)
at the New Technology Telescope \citep{White05}, and optical VLT spectroscopy
\citep{Halliday04,MilvangJensen08}, weak-lensing maps \citep{Clowe06}, galaxy
morphologies \citep{Desai07,Simard09}, fundamental plane parameters
\citep{Saglia10}, brightest cluster galaxy identifications \citep{Whiley08},
and MIPS-based star-formation rates \citep{Finn05,Finn10}. Wide-field imaging
in the mid-infrared with MIPS ($\sim50' \times 20'$ FOV) and ultraviolet
with \textit{Galaxy Evolution Explorer} ($\approx38'$ radius FOV) also exists for cluster subsets but does
not appear in this study.

%%%%%%%%%%%%%%%%%%%%%%%%%%%%%%%%%%%
\section{Wide-field Imaging Data}\label{sec:wfi}

We use wide-field ($\sim30' \times 30'$) imaging of our clusters to
identify targets for our LDP masks and to measure galaxy magnitudes and
colors, which are used for the redshift-fitting portion of the LDP
pipeline, as well as for characterizing the galaxies. Our photometry is measured from
$BVRIz$ images, with $VRI$ data from the Wide Field Imager (WFI) instrument on the
2.2m Max Planck Gesellschaft/European Southern Observatory (MPG/ESO) telescope
\citep{Baade99} and $Bz$ data from MOSAIC on the Cerro Tololo Inter-American
Observatory (CTIO) Blanco telescope, which have $34' \times 33'$ and $36' \times 36'$
FOVs, respectively. Not all clusters have been observed in all five bands. Our entire
sample has $VRI$ data, while some clusters appearing in \cite{Guennou10} have
either $B$ or $z$, or both (see Column~(9) of Table~1). Details on the imaging data
are given below.

\subsection{$V$-, $R$-, and $I$-band Data from WFI} %%%%%%%%%%%%%%%%%%%%%

We reduce the raw images using the techniques described by
\cite{Clowe01,Clowe02}, which involve bias-subtracting and flat-fielding
each chip separately and removing fringing in the $R$- and
$I$-band images. We calculate astrometric solutions for the images by
comparing the image centroids of U.S. Naval Observatory (USNO)
reference stars and use the utility {\tt imwcs} to write a new world
coordinate system (WCS) header based on those matches.\footnote{Originally written at the
University of Iowa, but since adapted and amplified by Jessica Mink at
the Smithsonian Astrophysical
  Observatory (http://tdc-www.harvard.edu/wcstools/imwcs/).} This
procedure results in an RMS position per star of $\approx0\farcs3$
relative to the USNO coordinates. For Cl$1354.2$--$1230$, this method
failed to converge, so we define the astrometry using SCAMP
\citep{Bertin06}. The astrometric precision is $\approx0\farcs5$ for
this field.

We create photometric catalogs using SExtractor \citep{Bertin96}. We
detect sources in the seeing-matched $R$-band image, requiring at
least 12 adjacent pixels containing flux $>5\sigma_{\rm RMS}$ above
the background. Photometry is performed in two-image mode for the
other bands. Given the wide FOV, we correct for Galactic extinction
differentially across the field. The color excess, which is
directly proportional to the extinction, across a given field varies by
$\approx 0.01$--$0.02$. We determine $E(B-V)$ at each
photometric source using the dust maps of \cite{Schlegel98}
and interpolate the extinction curve of \cite{Cardelli89} to the
effective wavelength of each bandpass to determine the extinction,
assuming $R_V=3.1$.

To match the point-spread functions (PSFs) among bands so that
aperture-matched magnitudes probe the same region of the galaxy, we smooth
the images with a gaussian kernel selected to match the image with the
largest seeing for that field (often the $V$ or $I$ band). The resultant
effective seeing is typically $1\farcs2$ (FWHM) for
the different fields, except Cl$1037.9$--$1243$, which has seeing
$\approx 2\arcsec$. For most clusters, the image quality or effective seeing
varies by less than $0\farcs1$ ($<0.5$~pixels) over the image; for Cl$1227.9$--$1138$,
Cl$1232.5$--$1250$, Cl$1353.0$--$1137$, Cl$1354.2$--$1230$, and Cl$1411.1$--$1148$
it varies $<0\farcs2$ ($<1$~pixel).

The WFI data were taken under nonphotometric conditions and therefore are poorly
calibrated. We adjusted the photometric zero-points (ZPs) for the $VRI$ data using
well-calibrated and deep VLT images taken as part of the original EDisCS program.
To determine the ZPs, we first cross-correlate stars from the WFI
images with those from the VLT using a $0\farcs5$ matching threshold,
resulting in $\sim20$--$100$ matches per field. We compare the
non-extinction-corrected VLT magnitudes of these matches with their counts
in $3\arcsec$-radius apertures on the WFI images and use linear regression to
calculate color terms ($a_{\rm \lambda}$) of the following form for each of the bands:
\begin{equation}
ZP_{\rm WFI,\lambda}=ZP_{\rm VLT,\lambda}+a_{\rm \lambda}({\rm color})_{\rm WFI,\lambda}.
\end{equation}
where $\rm (color)$ is $V-R$ for calibrating the $V$ and $R$ bands and $V-I$ for calibrating the $I$-band.
A first guess for the $V-R$ or $V-I$ color yields WFI ZPs with which we calculate new $VRI$
magnitudes using
\begin{equation}
m_{\rm WFI,\lambda}=-2.5\log_{10}({\rm counts}_{\rm WFI,\lambda})+ZP_{\rm WFI,\lambda}.
\end{equation}
These in turn give new $V-R$ or $V-I$ colors. This process is iterated until the WFI magnitude between successive iterations converged to $|\Delta m| \leq 0.01$~mag or until 20 iterations. Most sources converged within just a few iterations. The final calibration used the median color terms from all clusters, as this value is not expected to vary significantly between observations; however, the normalization of the conversion from the WFI to the FORS photometry was allowed to vary on a cluster-by-cluster basis. The uncertainties in $VRI$ ZPs are $0.12$, $0.07$, and $0.12$, respectively.

$R$-band imaging from the VLT was not available for Cl$1018.8$--$1211$, Cl$1059.2$--$1253$,
Cl$1232.5$--$1250$, Cl$1301.7$--$1139$, Cl$1353.0$--$1137$, Cl$1411.1$--$1148$, and
Cl$1420.3$--$1236$. In what directly follows, all bandpasses refer to the
VLT filters. We estimate $R$-band magnitudes from synthetic $R$-band
magnitudes that we obtain by fitting the $BVIK$ spectral energy
distributions (SEDs) with stellar templates from \cite{Hauschildt99}. We
also use this methodology on clusters with $R$-band photometry to
assess its accuracy, finding that the absolute value of the difference between
predicted and observed mean $R$-band magnitude is $<0.02$. For the seven clusters
without VLT $R$-band imaging, we use the $R$-band magnitudes derived in this way
for the iterated scheme described above when measuring ZPs.

To calculate colors, we use fixed apertures of $1\arcsec$ radii to
maximize signal-to-noise ratio; using a larger aperture introduces more noise
into our color measurement. We estimate the total magnitude using the
FLUX\_AUTO measurement from SExtractor, which fits sources with an
ellipse following the method of \cite{Kron80}.

Given the nonstandard method for calibrating our data, we further assessed the quality of our ZP estimates by fitting photometric redshifts using the EAZY \citep{Brammer08}.  In some clusters, comparison of the results from an initial pass of EAZY showed severe offsets between the photometric and VLT/FORS2 spectroscopic or preliminary LDP redshifts. The $B$- and $z$-band data had been taken in photometric conditions and were properly calibrated, so we assumed that these had appropriate ZPs.  Many of the redshifts were being skewed toward values that suggested that the $VRI$ filters were the source of the problem. This is unsurprising given the assumptions we made when calibrating the WFI data using the FORS photometry. To account for any offsets that may have been introduced in the $VRI$ photometry by this method, we explored whether small offsets in the photometry could improve the photometric redshift performance.  We created a grid of ZP offsets in the range $-0.2\leq \Delta m\leq 0.2$ and looked for the combination of $VRI$ magnitude offsets that minimized the quantity
\begin{equation}
z_q = \frac{\sum^N_{i=0}(z_{s,i} - z_{p,i})^2}{N}
\end{equation}
where $z_{s,i}$ is the $i$th spectroscopic or LDP redshift, $z_{p,i}$ is the $i$th photometric redshift, and $N$ is the total number of redshifts for a given cluster.  The minimization of this $z_q$ parameter was performed on a randomly selected subset of half the galaxies, and our shifts were then tested on the remaining half of the spectroscopic sample. These magnitude ZP corrections result in a median improvement over all clusters in the mean of $|z_s-z_p|$ of 0.02 and in the biweight midvariance of  $|z_s-z_p|$ of 0.015.  For some clusters the improvement in the mean of $|z_s-z_p|$ was by as much as 0.06 and in the biweight midvariance of 0.08.

We estimate photometric errors by placing $10^3$ background apertures on
each image with radii ranging from $1$ to $6$ pixels ($\approx0.2$--$3$\arcsec),
avoiding sources by using the SEGMENTATION output of SExtractor. We fit the RMS
fluctuation of counts in each aperture as a function of aperture size to determine
the error at $1\arcsec$ and, for the AUTO magnitudes, at the Kron radius. These
errors are typically $\approx0.01$--$0.02$ for most of the galaxies that appear
in this paper, although at the magnitude limit (see below), the $V$-, $R$-, and
$I$-band errors approach $0.03$, $0.02$, and $0.03$, respectively. Errors on $V-I$
color are $\approx0.1$ or less.

Comparison with the VLT magnitudes suggests an RMS precision that varies
with limiting magnitude, in that the RMS increases at faint magnitudes. If we
only include WFI galaxies with $R_{\rm AUTO}<23.3$ (our photometric
completeness limit; see below), the RMS precision is $0.12$, $0.07$,
and $0.12$ for the $V$, $R$ and $I$ total magnitudes, respectively. This
includes the Poisson uncertainty of counts in the aperture.

\subsection{$B$- and $z$-band Data from MOSAIC} %%%%%%%%%%%%%%%%%%%%%

We use $B$- and $z$-band data for nine of our clusters obtained by
\cite{Guennou10} with the CTIO Blanco telescope using MOSAIC. These
data were reduced with the MIDAS, SCAMP, and SWarp packages
\citep{Banse88,Bertin02,Bertin06}. Exposure times for the $B$ and
$z$ data are $11\times 600$s and $18\times800$s, respectively.
\cite{Guennou10} describe the data in more detail.

The $B$- and $z$-band ZPs have errors of $0.09$ and $0.07$, respectively
\citep{Guennou10}. The $B$-band ZPs were corrected for galactic extinction
using a single $E(B-V)$ value per field from the \cite{Schlegel98} dust maps; we
``de-correct'' the $B$-band ZPs so that we can correct each source for
extinction individually. After convolving the images to match the largest
seeing (often the $V$ or $I$ band), we applied the $B$- and $z$-band ZPs
to their corresponding photometric catalogs generated in two-image mode
based on detections in $R$ (see \S3.1) and correct for extinction
differentially using the method described above.

\subsection{Rest-frame Magnitudes and Colors}\label{sec:kcorr}

In \S\ref{sec:infall}, we use rest-frame absolute $B$-band magnitudes when selecting
galaxies for the analyses, and we also use rest-frame $U-B$ colors for
color magnitude diagrams (CMDs) and $B-V$ colors to calculate stellar masses in
\S\ref{sec:cmd}. These rest-frame magnitudes and colors are calculated using $Q=4$ LDP
redshifts (see \S\ref{sec:ldp}) with EAZY \citep{Brammer08}.  This code fits the BVRIz photometry (or subset thereof; see Table 1) using linear combinations of a set of theoretical templates that have been reduced to a subset of fig ``principal component templates'' as described in \citet{Brammer08}.  These are the same templates used to derive photometric redshifts.  In effect the templates serve to interpolate between the observed data points or to extrapolate beyond the bounds of the observed photometry.  For some of our lowest-redshift clusters with no $B$-band observations, our bluest observed filter is slightly redward of the redshift rest-frame $U$-band filter, though always by less than $500$~\AA\ for cluster galaxies.  Thus, EAZY must slightly extrapolate to measure the
$U-B$ color.  For these clusters we see no systematic offset in their colors compared to the rest of the systems.

\subsection{Photometric Completeness} %%%%%%%%%%%%%%%%%%%%%%%

\begin{figure}
\epsscale{1.0}
\plotone{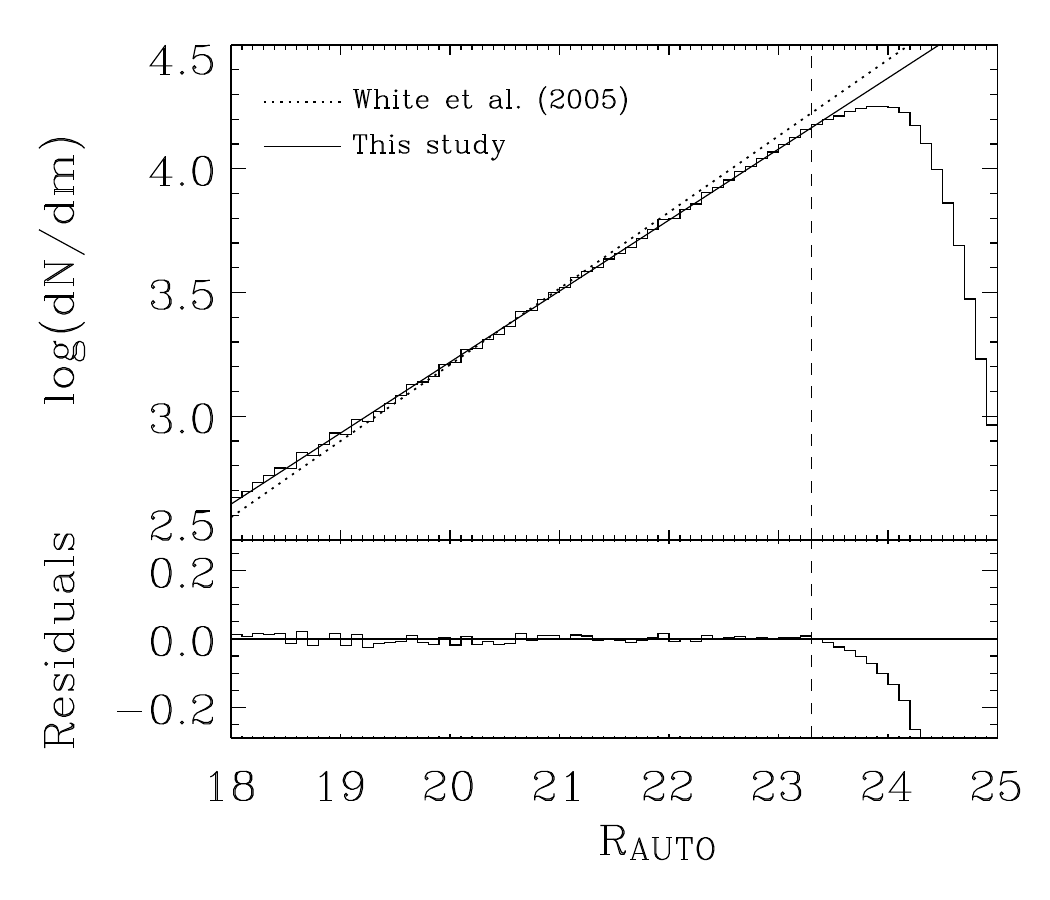}
%\figurenum{1}
\caption{\label{magcolor} ({\it Top panel}) Differential number counts
  of $R$-band detected sources per $0.1$-sized magnitude bin ($dN/dm$) as
  as function of $R_{\rm AUTO}$. We have not removed stars from the
  distribution. The solid line shows a fit to the distribution using a
  linear regression, while an estimate of $dN/dm$ for the core VLT
  photometry from \cite{White05} is shown as a dotted line. ({\it Bottom
  panel}) Residuals from the best fit show that $dN/dm$ follows a power law
  until $R_{\rm AUTO}=23.3$ (vertical dashed line), a clear sign of
  incompleteness beyond that magnitude.}
\end{figure}

We estimate our magnitude-limited completeness by examining the galaxy number counts as a function of magnitude (Figure~\ref{magcolor}). Differential number counts with magnitude ($\log~dN/dm$) follow a power-law distribution until the shape of the curve turns over once the catalog starts to become incomplete, with deeper catalogs turning over at fainter magnitudes \citep[e.g., Figure~1 of][although they applied aperture corrections to their magnitudes that result in a slightly different slope and sharper cutoff at the faint end]{White05}.  In Figure~\ref{magcolor}, we fit $\log~dN/dm$ using a linear regression and find that it follows a power law until $R_{\rm AUTO}\approx 24.1$, at which point the distribution turns over. The limits for individual fields range from 23.3 to 24.5 with a standard deviation of 0.3mag.  Only one field (Cl1420.3-1236) has a limit brighter than 23.9.  To be conservative, we choose a limit for the whole survey that corresponds to that cluster, and thus we are photometrically complete to $R_{\rm AUTO}<23.3$.

%%
%% Figure 1
%%

%%%%%%%%%%%%%%%%%%%%%%%%%%%%%%%%%%%
\section{LDP Spectroscopic Data}\label{sec:ldp} %%%%%%%%%%%

In this section, we present details on the LDP target selection, as well as
the redshift-fitting procedure and results.

\subsection{Target Selection}\label{sec:ldptargets} %%%%%%%%%%

We utilize the LDP and the IMACS camera on the Magellan I Baade 6.5m
telescope at Las Campanas Observatory. This instrument provides
spectra with a resolution of $\mathcal{R}=\lambda /
\Delta\lambda\approx 20$--$120$ from red to blue wavelengths, an improvement over
the resolution achieved with photometric redshifts ($\mathcal{R}\sim5$). The
corresponding redshift precision is also improved, as is the overall
accuracy. \cite{Coil11} present more details about the prism and
camera characteristics.

Of the original 20 EDisCS fields, four were not targeted with the LDP
and so do not appear in this paper. Cl1119$-$1129 and Cl1238$-$1144
do not have NIR data; the former contains a
$\sigma=166$~km~s$^{-1}$ cluster, while the latter has only four spectroscopic
redshifts at the cluster distance \citep{MilvangJensen08}. Cl1122$-$1136
does not contain a confirmed cluster, and Cl1202$-$1224 was not observed
owing to the limited telescope time available.

We obtained the LDP data during two observing runs, from
$7$--$9$~February 2008 and $27$--$30$~March 2009 (Table~2). Slit dimensions are
$1\arcsec\times0.8\arcsec$, compared to $1\arcsec\times1.6\arcsec$ for
the bulk of the PRIMUS survey \citep{Coil11}; this choice allows the placing
of $\approx1800$--$2800$ slits per mask. We chose exposure
times of $64\times60$~s per mask and used nod-and-shuffle mode to
improve sky subtraction.

We observed each field with two masks, except Cl$1232$--$1250$, which was
observed with three masks. Portions of each field are masked out owing to the
presence of bright stars. The FOV covers $\sim0.2$ square degrees around each
cluster, corresponding to clustercentric distances of $\sim6$--$8$~Mpc.
Because each mask in a given field has a different center, the final footprint
for each field has a nonregular shape.

There are $(1$--$2)\times10^4$ sources in our WFI catalog within each
LDP footprint. Of these, we target $\approx3000$--$5000$ objects per field with
the LDP ($\approx 20\%$ of potential targets, although the percentage ranges
among the fields from $15\%$ to $40\%$). Galaxies are targeted depending
on their $R$-band magnitude relative to the brightest cluster galaxy (BCG).
Priority $\#1$ targets have $R_{\rm AUTO,BCG}-1<R_{\rm AUTO}\ltsim23$. After
targeting those, additional slits (priority $\#2$) are placed on sources with
$19\ltsim R_{\rm AUTO}<R_{\rm AUTO,BCG}-1$. In cases where the
$R_{\rm BCG,AUTO}<20$, there are no priority $\#2$ targets. Finally, we place
slits on any galaxies that do meet these criteria but are capable of being
targeted in that mask; these ``filler" slits are $\sim 10\%$ of the total.
The $R$-band ranges of nonfiller targets per field appear in Table~3.

%%
%% TABLE 2
%%
\begin{deluxetable}{ccl}
\tablecolumns{3}
\tablenum{2}
\tablewidth{0pt}
\tabletypesize{\footnotesize}
\tablecaption{LDP Observing Log}
\tablehead{
\colhead{Run\tablenotemark{*}} &
\colhead{Cluster} &
\colhead{Seeing (arcsec)}
}
\startdata
1 & Cl$1040.7$--$1155$ & $0.7$ \\
1 & Cl$1054.4$--$1146$ & $2.2$ \\
1 & Cl$1054.7$--$1245$ & $0.6$--$1.1$ \\
1 & Cl$1103.7$--$1245$ & $0.6$--$1.0$ \\
1 & Cl$1216.8$--$1201$ & $0.7$ \\
1 & Cl$1227.9$--$1138$ & $0.5$--$0.9$ \\
2 & Cl$1018.8$--$1211$ & $0.6$--$0.7$ \\
2 & Cl$1037.9$--$1243$ & $0.5$--$1.0$ \\
2 & Cl$1059.2$--$1253$ & $0.5$--$0.6$ \\
2 & Cl$1138.2$--$1133$ & $0.5$--$0.6$ \\
2 & Cl$1232.5$--$1250$ & $0.4$--$0.6$ \\
2 & Cl$1301.7$--$1139$ & $0.5$--$0.7$ \\
2 & Cl$1353.0$--$1137$ & $0.4$--$0.6$ \\
2 & Cl$1354.2$--$1230$ & $0.4$--$0.6$ \\
2 & Cl$1411.1$--$1148$ & $0.4$--$0.7$ \\
2 & Cl$1420.3$--$1236$ & $0.5$--$0.7$ \\
\enddata
\tablenotetext{*}{Run 1 took place 6~Feb 2008 to 8~Feb 2008;
  Run 2 took place 27~Mar 2009 to 30~Mar 2009.}
\end{deluxetable}

%%
%% TABLE 3
%%
\begin{deluxetable}{lccc}
\tablecolumns{4}
\tablenum{3}
\tabletypesize{\footnotesize}
\tablewidth{0pt}
\tablecaption{Photometric Targeting Criteria\tablenotemark{*}}
\tablehead{
\colhead{Cluster} &
\colhead{$R_{\rm AUTO,BCG}$} &
\colhead{Priority \#1} &
\colhead{Priority \#2}
}
\startdata
Cl$1018.8$--$1211$ & $19.64$ & $18.64$--$23.13$ & \nodata \\
Cl$1037.9$--$1243$ & $19.79$ & $18.79$--$22.89$ & \nodata \\
Cl$1040.7$--$1155$ & $21.17$ & $20.17$--$22.99$ & $18.99$--$20.17$ \\
Cl$1054.4$--$1146$ & $21.20$ & $20.20$--$22.93$ & $18.93$--$20.20$ \\
Cl$1054.7$--$1245$ & $21.09$ & $20.09$--$22.97$ & $18.97$--$20.09$ \\
Cl$1059.2$--$1253$ & $19.20$ & $18.20$--$23.11$ & \nodata \\
Cl$1103.7$--$1245$ & $22.87$ & $21.87$--$22.99$ & $18.99$--$21.87$ \\
Cl$1138.2$--$1133$ & $20.03$ & $19.03$--$22.67$ & \nodata \\
Cl$1216.8$--$1201$ & $20.56$ & $19.56$--$23.07$ & $19.07$--$19.56$ \\
Cl$1227.9$--$1138$ & $21.06$ & $20.06$--$22.99$ & $18.99$--$20.06$ \\
Cl$1232.5$--$1250$ & $19.12$ & $18.12$--$23.05$ & \nodata \\
Cl$1301.7$--$1139$ & $19.56$ & $18.56$--$23.05$ & \nodata \\
Cl$1353.0$--$1137$ & $20.29$ & $19.29$--$23.08$ & $19.08$--$19.29$ \\
Cl$1354.2$--$1230$ & $21.27$ & $20.27$--$22.97$ & $18.97$--$20.27$ \\
Cl$1411.1$--$1148$ & $20.79$ & $19.79$--$23.08$ & \nodata \\
Cl$1420.3$--$1236$ & $20.09$ & $19.09$--$22.95$ & \nodata \\
\enddata
\tablenotetext{*}{All $R_{\rm AUTO}$ magnitudes in this table are from the photometry on hand when the data were taken. The photometry has since been revised with changes $<0.5$ mag and typical changes of $\approx 0.2$ mag.}
\tablecomments{Priority \#1 corresponds to $R_{\rm AUTO,BCG}-1<R_{\rm AUTO}\ltsim23$;
  Priority \#2 corresponds $19\ltsim R<R_{\rm AUTO,BCG}-1$, when $R_{\rm BCG,AUTO}>19$.}
\end{deluxetable}

The mean separation between adjacent slits for an individual mask is
$\approx20\arcsec$, with a minimum separation of $10\arcsec$.
However, multiplexing done with multiple masks per field increases the
sampling density, with a mean separation of $\approx15\arcsec$ and
$15$--$20\%$ of slits separated by $<10\arcsec$ (with the closest
pairs $\approx 1\arcsec$ apart).

\subsection{LDP Redshifts}\label{sec:ldpz} %%%%%%%%%%%%%%%%%%%%%%

\begin{figure}
\epsscale{1.0}
\plotone{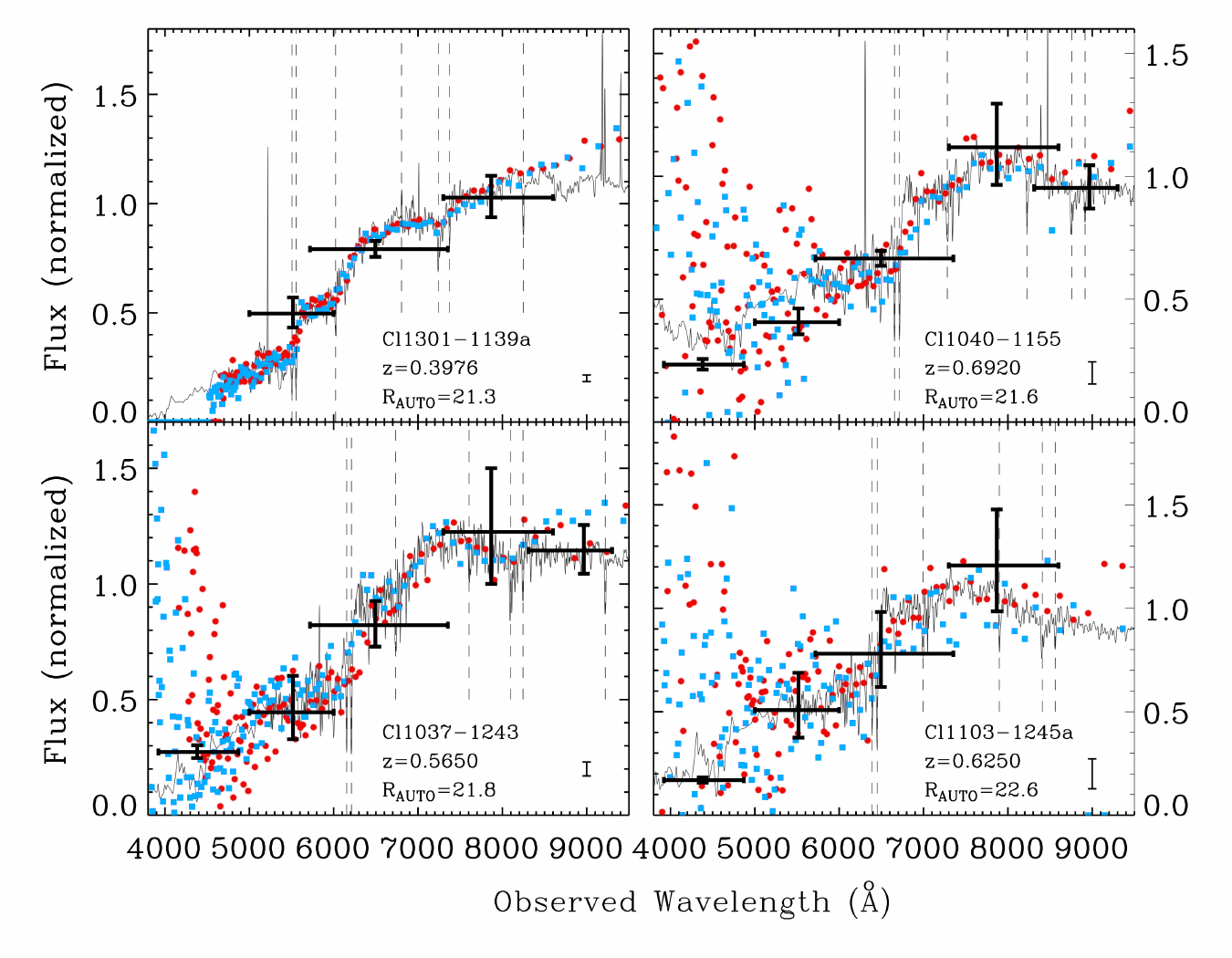}
%\figurenum{2}
\caption{\label{egspec} Sample LDP spectra with $Q=4$ for galaxies in four
  clusters, spanning $21.3<R_{\rm AUTO}<22.6$ and $<0.4<z<0.7$. Cyan squares and
  red circles are the LDP spectra values in each slit of the nod and shuffle, while
  the continua spectra shown are those of the best-fit templates. The spectra have
  been normalized to unity at $6800$\AA. Typical errors on the LDP data are shown
  at the bottom right of each panel. Also plotted are magnitudes ($VRI$, $BVRIz$,
  $BVRIz$, and $BVRI$, respectively), and prominent spectral features are shown as
  vertical dashed lines. The LDP redshifts agree with those measured from FORS2 to
  $\delta z=0.0024$, $-0.0132$, $-0.0139$, and $0.0012$.}
\end{figure}

The PRIMUS reduction pipeline simultaneously fits the spectral and
photometric data to a set of galaxy templates at different redshifts and
calculates a best-fit $\chi^2$ value at each redshift \citep{Cool13}.
While the relative astronometry of our observing masks was accurate, they
were mildly offset in absolute astronometry. We calculated a new astrometric
solution for the WFI imaging after the LDP slit positions were determined
by cross-correlating the WFI catalog with the slits using a $1\arcsec$ matching
threshold. The pipeline treats photometric data similarly to 1 pixel of the
spectrum. The $\chi^2$ fit is determined primarily from the LDP data, but the
photometric data help distinguish between redshift solutions. On the basis of
the $\chi^2$ distribution, both a best-fit redshift and a redshift confidence
parameter, $Q$ are calculated. The confidence parameter is assigned 
using the ratio of the width of the primary peak in the $P(z)$ distribution and
the goodness of fit between the first and second peaks. This ratio is then used to
assign an integer confidence parameter between $2$ and $4$, with $Q=4$ objects typically 
having a narrow primary peak compared to other features in the $P(z)$ distribution. 
Further details on the redshift-fitting procedure appear in \cite{Cool13}.
Example spectra of four cluster galaxies appear in Figure~\ref{egspec}. 

%%
%% Figure 2
%%

For objects with more than one redshift measured (given the multiple masks),
we take the redshift with the higher-quality flag, $Q$. In cases where there
are multiple redshifts with the same $Q$, we randomly select one. This is done to
avoid averaging significantly discrepant redshifts when they exist (see Figure~\ref{zvz}
for outlier rates).

We define cluster membership as galaxies that have a $Q=4$ best-fit redshift
within $\pm0.02$ ($\pm 6000$~km~s$^{-1}$) of the cluster redshift, which is
approximately three times the accuracy of the LDP redshifts (see
\S\ref{sec:accuracy}). We choose a fixed cut in redshift, rather than a multiple
of the cluster $\sigma$, because the LDP uncertainty is larger than any velocity
dispersion in our sample. We do not make any spatial cuts because we are interested
in galaxies at large clustercentric radii. This selection results in 1763
galaxies that we place in the cluster environment.

A summary of the number of LDP targets, redshifts, and cluster members
is presented in Table~4. Two of the LDP-observed clusters do not appear in this
study. Cl$1103.7$--$1245$, at $z=0.95$, with only three LDP-selected cluster members, 
does not have enough cluster members for a meaningful analysis. Cl$1138.2$--$1133$a,
at $z=0.4548$, lies too close to the redshift of Cl$1138.2$--$1133$, $z=0.4796$, to
distinguish between members using the $z_{\rm clus}\pm0.02$ selection. We therefore
include the latter cluster, with the caveat that some contamination may come from
galaxies belonging to the former. We note that Cl$1138.2$--$1133$ is not a
significant outlier in any of the analyses that follow.

%%
%% TABLE 4
%%
\begin{deluxetable}{lrrrrr}
\tablecolumns{6}
\tablenum{4}
\tabletypesize{\footnotesize}
\tablewidth{0pt}
\tablecaption{LDP Information}
\tablehead{
\colhead{Cluster} &
\colhead{$N_{\rm phot}$} &
\colhead{$N_{\rm targets}$} &
\colhead{$N_{\rm LDP}$} &
\colhead{$N_{Q=4}$} &
\colhead{$N_{\rm memb}$} \\
\colhead{(1)} &
\colhead{(2)} &
\colhead{(3)} &
\colhead{(4)} &
\colhead{(5)} &
\colhead{(6)}
}
\startdata
cl1018.8-1211 & 5349 & 1645 & 1425 & 781 & 86 \\
cl1037.9-1243 & 10380 & 2231 & 1977 & 1194 & 47 \\
cl1037.9-1243a & \nodata & \nodata & \nodata & \nodata & 189 \\
cl1040.7-1155 & 9597 & 2647 & 2218 & 962 & 31 \\
cl1054.4-1146 & 6297 & 2710 & 2337 & 1174 & 45 \\
cl1054.7-1245 & 9710 & 2509 & 2070 & 1076 & 89 \\
cl1059.2-1253 & 9341 & 2275 & 2135 & 1458 & 154 \\
cl1103.7-1245a & 9392 & 2561 & 1479 & 584 & 23 \\
cl1103.7-1245b & \nodata & \nodata & \nodata & \nodata & 16 \\
cl1138.2-1133 & 6088 & 1530 & 1406 & 1143 & 84 \\
cl1216.8-1201 & 9435 & 2557 & 2086 & 1022 & 44 \\
cl1227.9-1138 & 9048 & 2612 & 1983 & 1238 & 86 \\
cl1227.9-1138a & \nodata & \nodata & \nodata & \nodata & 105 \\
cl1232.5-1250 & 9947 & 2455 & 2227 & 1597 & 166 \\
cl1301.7-1139 & 9426 & 1683 & 1497 & 1085 & 131 \\
cl1301.7-1139a & \nodata & \nodata & \nodata & \nodata & 158 \\
cl1353.0-1137 & 11425 & 2222 & 1951 & 1242 & 39 \\
cl1354.2-1230 & 9483 & 2269 & 2040 & 1468 & 38 \\
cl1354.2-1230a & \nodata & \nodata & \nodata & \nodata & 80 \\
cl1411.1-1148 & 10485 & 1897 & 1674 & 1157 & 76 \\
cl1420.3-1236 & 10103 & 1477 & 1318 & 841 & 76 \\
\hline
Total & 145506 & 35280 & 29823 & 18022 & 1763
\enddata
\tablecomments{All numbers only include galaxies brighter
than $R<22.9$, our spectroscopic completeness limit. Numbers for
columns $2$--$5$ for serendipitously discovered clusters are
suppressed as they are in the same field as the primary cluster.
(1) cluster name; (2) number of photometric
sources; (3) number of LDP targets; (4) number of successfully
extracted LDP spectra; (5) number of $Q=4$ LDP spectra ;
(6) number of cluster members (defined by $z_{\rm clus}\pm0.02$).}
\end{deluxetable}

\subsubsection{LDP Redshift Accuracy}\label{sec:accuracy}
\begin{figure}
\epsscale{0.95}
\plotone{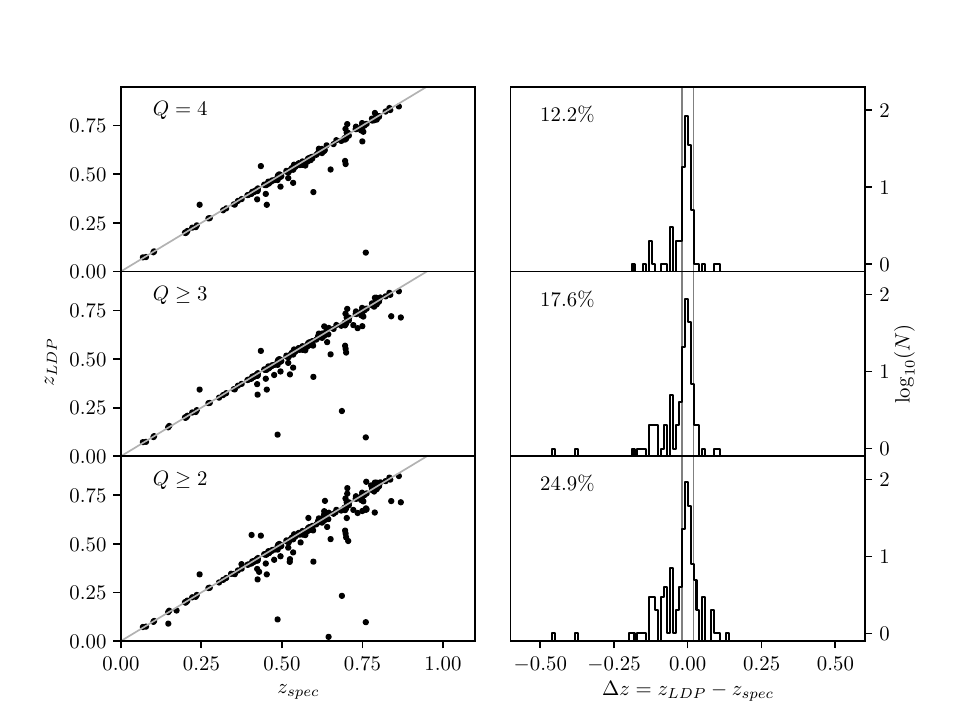}
%\figurenum{3}
\caption{\label{zvz} ({\it Left panels}) Comparison of LDP-derived redshifts
  ($z_{\rm LDP}$) with spectroscopic redshifts ($z_{\rm SPEC}$). The LDP redshifts
  are split showing different cuts in the quality flag, $Q$. The
  outliers in $z_{\rm LDP}$ systematically underestimate the redshift. 
  ({\it Right panels}) Histograms of the residuals from
  the left panels. Vertical lines show $\pm0.02$, which is the size of the
  redshift interval used in selecting cluster galaxies. Percentages
  in the upper left corner show the fraction of outliers outside this
  interval. The accuracy with the $z_{\rm LDP}$ ($\sigma=0.007$ for $Q=4$) is an
  order-of-magnitude improvement over the photometric redshifts.}
\end{figure}

We assess the accuracy of our LDP-derived redshifts ($z_{\rm LDP}$)
by comparing them to the subset of 427 galaxies also observed with
VLT/FORS2 ($z_{\rm SPEC}$) over a wide range of redshifts and
with photometric redshifts ($z_{\rm PHOT}$) calculated in \cite{Pello09} from
$BVIK$, $BVIJK$, and $VRIJK$ imaging of the cluster cores; these filter combinations
were chosen based on the initial redshift estimate of the cluster. We match galaxies
within $1\arcsec$, and show the results of these comparisons in Figure~\ref{zvz}. We
only consider galaxies with $z_{\rm LDP}<0.85$, which is just above our highest-redshift
cluster (Cl$1216.8-1201$ at $z=0.79$); considering the full
range of redshifts that PRIMUS fits (out to $z=1.2$) affects neither
the accuracy nor the outlier rate significantly.

The LDP-derived redshifts are more precise than the photometric
redshifts by an order of magnitude, their RMS being 
$\sigma(|z_{\rm LDP}-z_{\rm SPEC}|)=0.007$ for $Q = 4$
data, compared to $0.08$ for the photometric
redshifts. The outlier rates of LDP-derived redshifts, defined as
$|z_{\rm LDP}-z_{\rm SPEC}|>0.02$, depend on the quality cut and
range from $25\%$ ($Q\geq2$) to $18\%$ ($Q\geq3$) to $12\%$
($Q=4$).

%%
%% Figure 3
%%

\begin{figure}
\epsscale{1.0}
\plotone{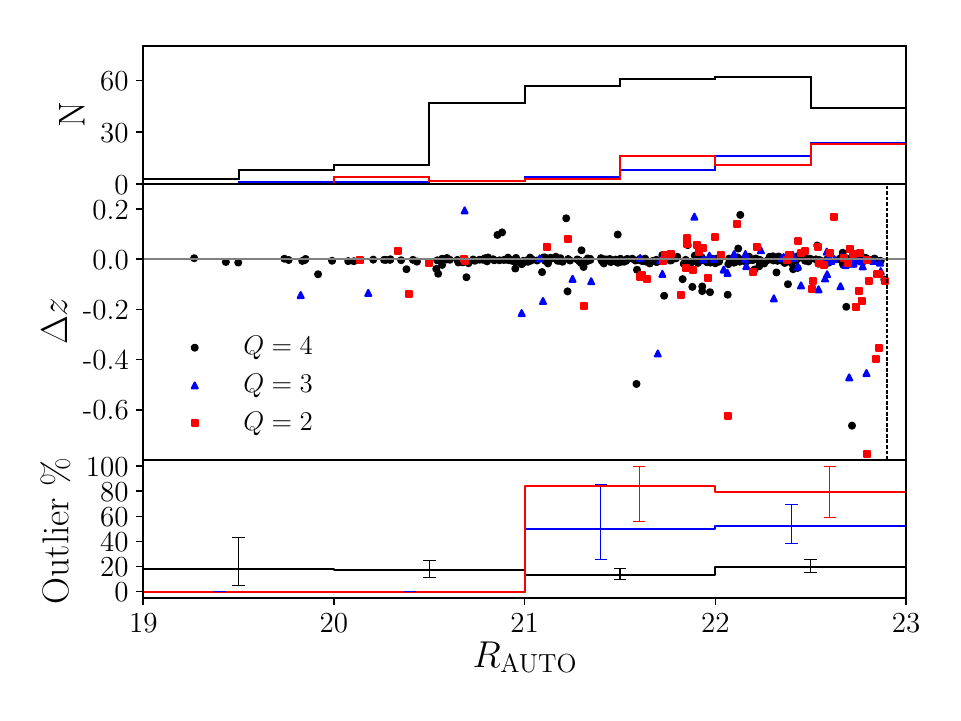}
%\figurenum{4}
\caption{\label{dzmag} ({\it Central panel}) Residuals between LDP and
  FORS2 redshifts as a function of $R$-band magnitude for $Q=4$ (black circles),
  $Q=3$ (blue triangles), and $Q=2$ (red squares) redshifts.
  ({\it Top panel}) Histograms of $R$-band magnitude and
  $z_{\rm LDP}-z_{\rm SPEC}$ residuals.
  ({\it Bottom panel}) Outlier rate ($|\Delta z| >0.02$) as a function of $R$-band
  magnitude. The outlier rate is approximately flat for a given $Q$
  flag for bins containing more than five objects. The vertical dotted line is our spectroscopic completeness limit.}
\end{figure}

\begin{figure}
\epsscale{1.0}
\plotone{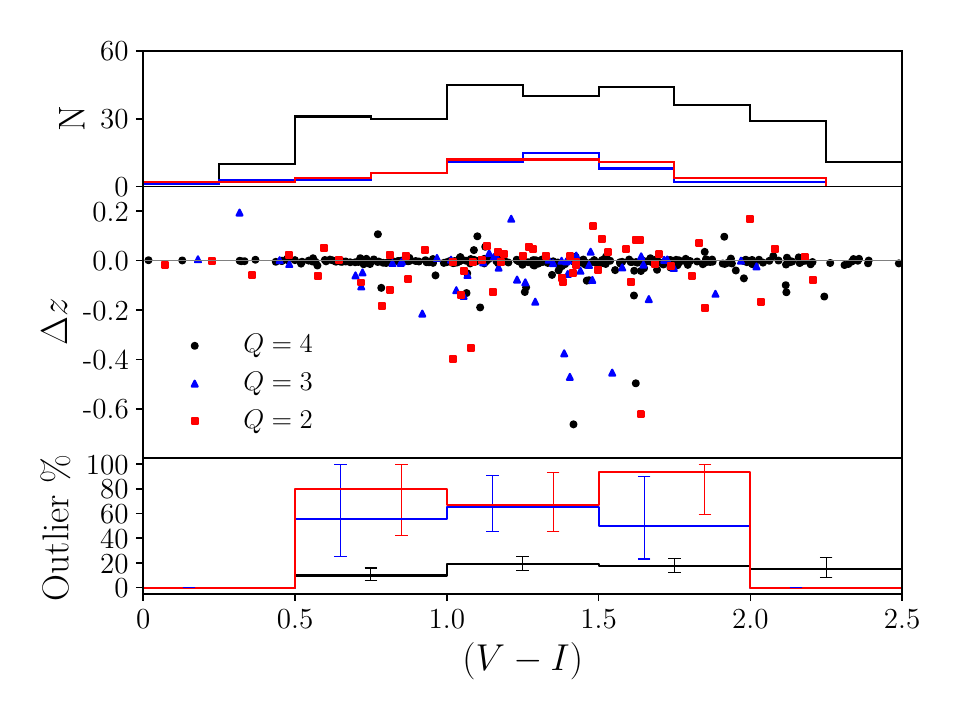}
%\figurenum{5}
\caption{\label{dzcol} ({\it Central panel}) Residuals between LDP and
  FORS2 redshifts as a function of $V-I$ color. Symbols are the same
  as Figure~\ref{dzmag}. ({\it Top panel}) Histogram of $V-I$ color.
  ({\it Bottom panel}) Outlier fraction as a function of $V-I$ color.
  The outlier rate is approximately flat for a given $Q$ flag 
  for bins containing more than five objects.}
\end{figure}

In Figure~\ref{dzmag} we plot the LDP redshift accuracy and outlier
rate dependence on $R$-band magnitude. The precision between the LDP redshifts
and true spectroscopic redshifts is constant, even at faint magnitudes for the high-quality
$Q=4$ spectra. Above $R_{\rm AUTO}>21$, we see a large
increase in the number of $Q=2$ and $3$ spectra, which results in a significant increase in outlier rate.
The outlier rate is constant with magnitude for spectra of a given $Q$ value. 
We do not find any significant dependence of the redshift accuracy on $V-I$ color 
(Figure~\ref{dzcol}). The objects with $Q=4$ redshifts make up most of our 
redshift catalog and have a stable outlier fraction of $12\%$.

%%
%% Figure 4
%%

%%
%% Figure 5
%%

From Figures~\ref{zvz} and \ref{dzmag}, it is apparent that the
outliers in LDP redshift are skewed toward lower-redshift values (i.e.,
the LDP fits a lower redshift than the ``true" one). Although the
outlier rate does not depend on galaxy color, we find a significant
dependence on [OII] emission. Being relatively close to the Balmer
break, the blending of the two spectral features ``drags" the
break to a lower redshift, consistent with the bias evident in
Figure~\ref{zvz}. When we consider $Q=4$ LDP spectra, we find that galaxies
with [OII] equivalent widths (EWs) of $\sim5$\AA\ have an outlier rate of
$35\%$, almost triple the outlier rate for all galaxies ($12\%$).
Moreover, at both higher and lower EWs, the outlier rate drops. This can
be understood as galaxies with weaker [OII] emission not suffering from
this blend ``dragging" the Balmer break to a lower redshift, while
galaxies with stronger [OII] emission have lines that become the dominant
redshift feature. Therefore, we are more likely to miss cluster galaxies that
have modest [OII] emission at an outlier rate twice as high as for the full
galaxy population, or $\sim4\%$ of the cluster sample, and have an some
enhanced contamination from field galaxies with modest [OII] emission at
higher redshift.

\subsubsection{Spectroscopic Completeness}\label{sec:speccomplete} %%%%%
\begin{figure}
\epsscale{0.9}
\plotone{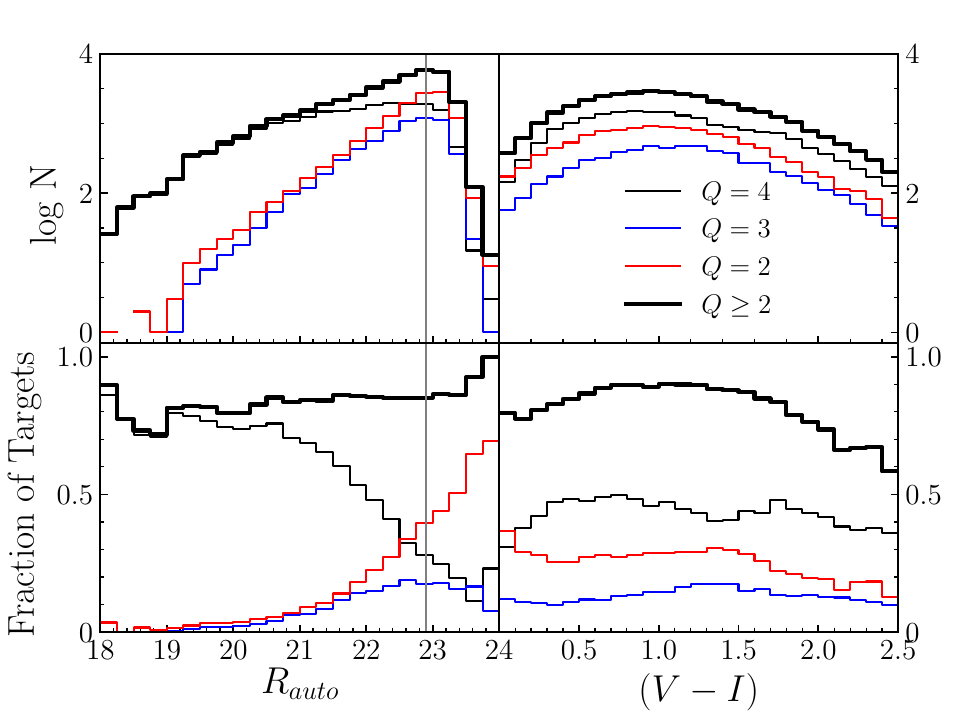}
%\figurenum{6}
\caption{\label{slitmag} ({\it Top left panel}) Histogram of $R$-band magnitudes
  for LDP-targeted galaxies with bin sizes of $0.1$ for different values of $Q$. The
  spectroscopic completeness limit at $R_{\rm AUTO}=22.9$ is shown as a vertical line.
  ({\it Bottom left panel}) The fraction of targets with a successfully extracted
  spectrum, i.e. number of spectra divided by number of slits, as a function of magnitude.
  At bright magnitudes, most of the successfully extracted spectra have $Q=4$.
  ({\it Right panels}) Same as the left panels, only as a function of $(V-I)$ color.}
\end{figure}

Figure~\ref{slitmag} shows the distribution of $R_{\rm AUTO}$
(analogous to Figure~\ref{magcolor}) for the LDP targets with
successfully measured redshifts. The different curves show the
distributions for $Q=2$, $3$, and $4$ redshifts. At the brightest
magnitudes, the vast majority of redshifts have a secure $Q=4$
flag. However, at $R_{\rm AUTO}\gtsim 20$, redshifts with lower $Q$
flags begin to appear in significant numbers. For all $Q$ values, the
distribution turns over before $R_{\rm AUTO}\sim23$; our photometric
catalog is therefore complete to fainter magnitudes than our
spectroscopic one. The full distribution (including all $Q$ values)
departs from a power law at $R_{\rm AUTO}\approx 22.9$, which we take
as the estimate of our spectroscopic completeness.

\subsubsection{Radial Completeness}\label{radcomplete} %%%%%%%%%%%%%%%%%%%%%%%
\begin{figure}
\epsscale{1.0}
\plotone{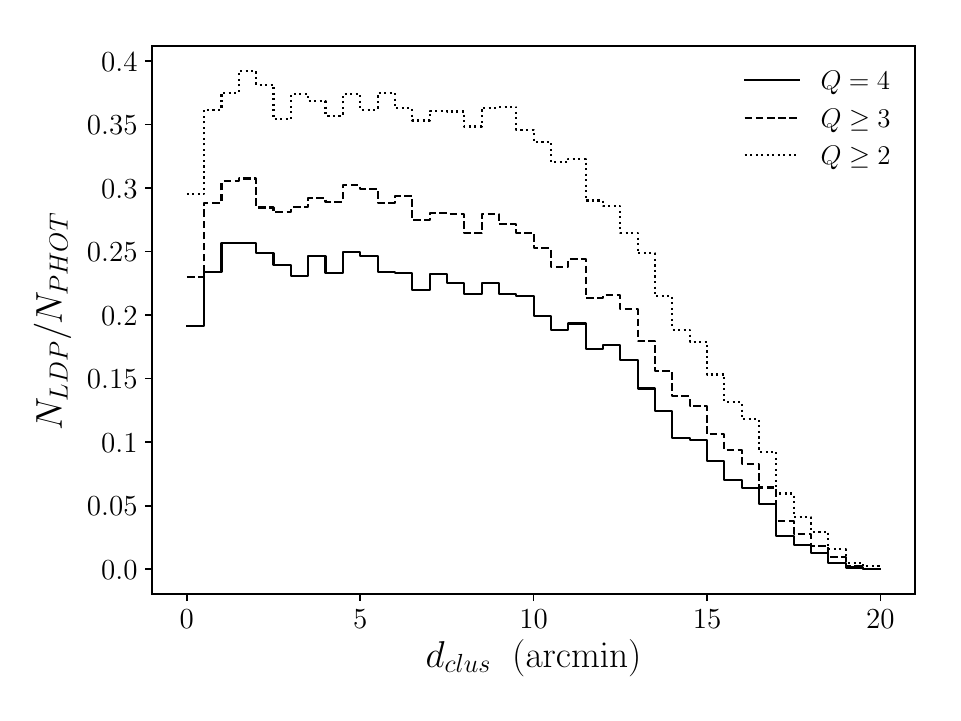}
%\figurenum{7}
\caption{\label{slitrad} Fraction of photometric sources with LDP
  redshifts ($N_{\rm LDP}/N_{\rm PHOT}$) as a function of
  clustercentric angular distance ($d_{\rm clus}$). A distance of $10\arcmin$
  corresponds to $\sim 4$~Mpc, at $z=0.6$.
  Bin sizes are $0.5\arcmin$, and solid, dashed, and
  dotted lines include redshifts with $Q=4$, $Q\ge3$, and $Q\ge2$,
  respectively.}
\end{figure}

We also quantify the percentage of successfully measured redshifts as a
function of clustercentric distance. We consider the fraction of
successfully measured redshifts relative to the number of photometric sources,
restricting both to galaxies brighter than our spectroscopic completeness
($R_{\rm AUTO}<22.9$), as a function of angular distance from the
cluster ($d_{\rm clus}$). Figure~\ref{slitrad} shows that the percentage of
targets with measured redshifts is $\sim 25$--35$\%$, depending on $Q$-cut, out to
$\sim10'$. Converting this to a physical distance for our typical clusters puts
the drop-off at $\sim4$~Mpc. The percentage then drops off, as most fields have
only one mask coverage at these radii, and approaches zero smoothly rather than
abruptly because the offset placement of masks leads to an edge that is not
spherically symmetric about the cluster.

%%
%% Figure 6
%%

%%
%% Figure 7
%%

Given the surface density of slits on the sky, we are also relatively
insensitive to close pairs. While multiplexing with two to three masks per
field allows us to measure redshifts for galaxies $\sim1\arcsec$
apart that are not near the edges of the footprint, this distance is small
compared to the average separation between adjacent slits
($\approx 20\arcsec$). Only $\approx10\%$ of slits have
separations of $10\arcsec$ or less. Compared to frequency of slits with
separations of $10$--$30\arcsec$, the frequency of slits separations
$<10\arcsec$ is only $\approx 40\%$ of that value. This affects our ability to
find cluster galaxies within the cluster core, where the galaxy surface
densities are higher (see \S\ref{sec:infall}).

%%%%%%%%%%%%%%%%%%%%%%%%%%%%%%%%%%%%

\section{Cluster Infall Regions}\label{sec:infall}
We use the theory of secondary infall \citep{Fillmore84,Bertschinger85,White92}
to estimate (1) the masses that our clusters are expected to accrete by $z=0$
($M_{\rm infall}$) and (2) the projected radii at the cluster redshifts that
encloses $M_{\rm infall}$ ($R_{\rm infall}$). Using the latter with the LDP data,
we calculate the number of galaxies in the infall region ($N_{\rm infall}$). In
\S\ref{sec:models}, we calculate the expected evolution of the clusters in terms
of mass and compare to semianalytic models. In \S\ref{sec:n_infall}, we focus on
the dependence of $N_{\rm infall}$ on cluster velocity dispersion and quantify
its scatter. In \S\ref{sec:cmd}, we examine the quiescent fraction of galaxies
in different environments, using the red sequence galaxy fraction as a proxy, and
quantify the amount of clustering in the infall regions.

\subsection{The Secondary Infall Model}\label{sec:models}

While previous studies of cluster infall regions have used the caustic technique
\citep[e.g.,][]{Gellar99,Rines03,Rines06,Serra11,Owers17}, which identifies curves in
galaxy position--radial velocity phase space that encompass those galaxies that are
gravitationally bound to the cluster, such an analysis requires redshift
accuracy greater than that of the LDP. Alternatively, we identify the infall
regions of our clusters using the theory of secondary infall to define projected
radii that encompass the infall region.

The secondary infall model describes how shells of mass centered on a cosmic
perturbation evolve over time. The shells begin by expanding outward, until a time
$t_{\rm turn}$ when they turn around owing to the pull of gravity. The shells do not
cross during this time, and there is a critical mass, $M_*$, enclosed by the
shell that is marginally bound. All shells enclosing a mass less than $M_*$
eventually turn around at different times and collapse, while shells at larger
radii continue to expand forever. We follow the equations of \cite{White92}, who
assume an open universe with $\Omega_\Lambda=0$. Keeping $\Omega_\Lambda=0$ when
calculating $M_{\rm infall}$ and $R_{\rm infall}$ does not significantly affect
the analysis given the physical scales involved \citep[e.g.,][]{DelPopolo12}, as
we also confirm with our own comparison to cosmological simulations discussed
below. For determining global quantities (e.g., connecting a time with a
redshift and age), we continue to use $\Omega_\Lambda=0.7$.

Our aim is to compare the ``mass" of the cluster at the observed time to that at the current time. We approximate the cluster's mass by measuring the mass of material that has reached the cluster center at least once by the time of interest. For the case of the mass at the observed time, that time corresponds to the age of the universe at the observed redshift.  To calculate the mass that has reached the center, we use the equations of secondary infall and calculate the mass enclosed within the turnaround radius ($R_{\rm turn}$) at half the age of interest. For the observed clusters we approximate the enclosed mass using $M_{200}$ (Table~1) and then calculate $M_*$ in the following equation when we set the turnaround time $t_{\rm turn}$ to half the age of the universe
\begin{equation}
  t_{\rm turn}(M_{\rm enc})=\frac{\pi}{2}\frac{\Omega}{H(1-\Omega)^{3/2}}\left[\left(\frac{M_*}{M_{\rm enc}}\right)^{2/3}-1\right]^{-3/2},
  \label{eq:turnaround}
\end{equation}
%%%%%%%%%%%%%%%%%
Once $M_*$ is determined, we use Equation~\ref{eq:turnaround} again, with $t_{\rm turn}$ equal to half the present age of the universe, to determine the enclosed mass or approximate $M_{200}$ at $z = 0$. The ratio of the mass at the current time to that at the observed time is presented in Table~5. These quantities are calculated separately for each cluster, which is why two clusters with nearly the same velocity dispersion (e.g., clusters 3 and 6 in Table~5) can have quite different mass ratios. 
$R_{\rm turn}$ for this shell is related to $t_{\rm turn}$ through the simple equation for a
free-falling test particle,
\begin{equation}
  t_{\rm turn}(M_{\rm 200,z=0})=\frac{\pi}{2}\sqrt{\frac{R_{\rm turn}^3}{2GM_{\rm 200,z=0}}}.
\end{equation}
%%%%%%%%%%%%%%%%%
While we solve for $R_{\rm turn}$ for each cluster, which turns around when the
universe is half its present age, what we are truly interested in is the location of these
shells at $z=z_{\rm clus}$. Therefore, we use the equation of motion for a uniform mass
shell to evolve $R_{\rm turn}$ to $z=z_{\rm clus}$. These radial distances are the infall
radii ($R_{\rm infall}$), the outer boundaries of the relevant infall regions for $z=0$
observations.

In defining radial distances, we center on the location of the BCG. However, the BCG may be
offset from the distribution of mass, which would
affect the definition of the infall region. We estimate the magnitude of these offsets
from Figure~6 of \cite{White05}, which marks the BCG position relative to adaptively
smoothed contours of cluster galaxy surface density. The offsets are $\ltsim10\%$ of
$R_{\rm infall}$ for all clusters except Cl$1037$--$1243$, whose BCG is
offset by $\sim25\%$ of $R_{\rm infall}$. These values are larger than the
typical offsets found at $z\approx0.5$ by \cite{Zitrin12}, but they find that the offsets
are positively correlated with redshift and our clusters lie at higher $z$ than their sample.
In addition, the galaxy distributions from \cite{White05} include galaxies with photometric redshifts
consistent with being close to the cluster redshift and therefore include a non-negligible number of
interlopers that make the centering less precise. If we define the center to be the mean R.A. and decl.
of the VLT/FORS2 spectroscopic sources, then $41$ galaxies ($12\%$ of the infalling population)
are either removed or added by the new definition. However, the total number of infalling
galaxies changes by less than $2\%$ (because some are added, while some are removed from the
infall region), and the fraction of red galaxies (\S\ref{sec:cmd}) changes by only $0.4\%$.
As a final check, we randomly apply offsets of $\sim0.2$~Mpc in various directions from the
BCGs and find that similar numbers of galaxies are affected by the redefined infall regions. We
conclude that reasonable uncertainties in the centering of the clusters do not strongly impact
our results or conclusions.

In Table~5, we present the results of the models for our clusters. In addition
to $R_{\rm infall}$, we calculate the predicted mass (and corresponding velocity
dispersion) at $z=0$ for our sample. The infall radii range from $1.2$ to $6.7$~Mpc; the ratio of
$R_{\rm infall}$ to $R_{\rm 200}$ is set entirely by the redshift of the cluster, in
that higher-$z$ clusters have larger ratios, and range from $3.0$ to $4.2 R_{\rm 200}$. These are smaller
than the ``turnaround radii" calculated in other studies of cluster infall regions, such as those found in
\citet[][$\approx 4.75 R_{\rm 200}$]{Rines06}. However, their
definition of infall region includes all galaxies that, with a velocity less than the cluster
escape velocity, will eventually become incorporated into the cluster given enough time,
while our model only includes galaxies that could have reached the center of the cluster
by $z=0$.

We compare our infall radii to models using the Millennium Simulation \citep{Springel05},
estimating the infall radii in the simulation using the fraction of galaxies
at a given clustercentric distance that come to lie within $R_{\rm 200}$
at $z=0$. We considered $174$ $\sim10^{14}$\msun\ halos at $z\sim0.6$, which have
$R_{\rm infall}\approx 3.5R_{\rm 200}$ according to our analytic modeling. We find that
$67\% \pm7\%$ of the galaxies within $3.5R_{\rm 200}$ at $z=0.62$ ultimately lie within
the virial radius of the descendant halo (the errors are the $15$th--$85$th percentiles). Because we
assume that $100\%$ of galaxies within $R_{\rm infall}$ become ``cluster galaxies'' of the
descendant halo, our prediction is good to $\approx 33\%$. The difference arises from a
variety of effects, including the assumption of spherical symmetry, the definition of an infalling
galaxy as one that reaches $R=0$ at $z=0$, cluster galaxy dynamics such as merging and tidal stripping, 
and the presence of ``backsplash'' galaxies, which pass
through the cluster core and then continue out to radii larger than $R_{\rm 200}$ \citep{Balogh00}.
We also acknowledge that galaxies from $R>R_{\rm infall}$ may be measured within the virial radius
at $z=0$, but we expect that this effect is small and subdominant to the other sources of uncertainty.

Our models predict an increase in cluster mass of $26$--$57\%$ from the observed
epoch to the current one (Column~(7) of Table~5). Velocity dispersion increases
as
\begin{equation}
  \sigma\propto M^{1/3} [\Omega_\Lambda + \Omega_0 (1+z)^3]^{1/6},
\end{equation}
which means that the mean growth corresponds to $\sigma$ increasing by $\approx15$--$25\%$. The
predicted $\sigma$ at $z=0$ agrees to within $\sim10\%$ of the predictions at a given mass
that \cite{Poggianti06} computed for $z=0.6$ clusters by combining the high-resolution
$N$-body simulations of \cite{Wechsler02} with cluster concentration parameters from
\cite{Bullock01}.

%%
%% TABLE 5
%%
\begin{deluxetable}{rlccccccc}
\tablecolumns{9}
\tablenum{5}
\tabletypesize{\scriptsize}
\tablewidth{0pt}
\tablecaption{Mass Infall Model Results}
\tablehead{
\colhead{Field} &
\colhead{Cluster} &
\colhead{$z$} &
\colhead{$R_{\rm infall}$} &
\colhead{$\frac{R_{\rm infall}}{R_{\rm 200}}$} &
\colhead{$M_{\rm 200,z=0}$} &
\colhead{$\frac{M_{\rm 200,z=0}}{M_{\rm 200}}$} &
\colhead{$\sigma_{\rm z=0}$} &
\colhead{$\frac{\sigma_{\rm z=0}}{\sigma}$} \\
%\colhead{$\alpha$} \\
\colhead{(1)} &
\colhead{(2)} &
\colhead{(3)} &
\colhead{(4)} &
\colhead{(5)} &
\colhead{(6)} &
\colhead{(7)} &
\colhead{(8)} &
\colhead{(9)}
}
\startdata
 1 & Cl$1018.8$--$1211$ & 0.4734 & 3.06 & 3.28 & 2.01(14) & 1.32 &  533 & 1.10 \\
 2 & Cl$1037.9$--$1243$ & 0.5783 & 2.07 & 3.60 & 5.67(13) & 1.40 &  357 & 1.12 \\
 3 & Cl$1037.9$--$1243$a& 0.4252 & 3.31 & 3.12 & 2.72(14) & 1.28 &  584 & 1.09 \\
 4 & Cl$1040.7$--$1155$ & 0.7043 & 2.75 & 3.94 & 1.27(14) & 1.50 &  478 & 1.14 \\
 5 & Cl$1054.4$--$1146$ & 0.6972 & 3.87 & 3.92 & 3.55(14) & 1.49 &  673 & 1.14 \\
 6 & Cl$1054.7$--$1245$ & 0.7498 & 3.32 & 4.05 & 2.21(14) & 1.53 &  581 & 1.15 \\
 7 & Cl$1059.2$--$1253$ & 0.4564 & 3.19 & 3.23 & 2.33(14) & 1.31 &  558 & 1.09 \\
 8 & Cl$1103.7$--$1245$a& 0.6261 & 2.20 & 3.73 & 6.61(13) & 1.43 &  379 & 1.13 \\
 9 & Cl$1103.7$--$1245$b& 0.7031 & 1.66 & 3.93 & 2.78(13) & 1.49 &  288 & 1.14 \\
10 & Cl$1138.2$--$1133$ & 0.4796 & 4.62 & 3.30 & 6.88(14) & 1.32 &  804 & 1.10 \\
11 & Cl$1216.8$--$1201$ & 0.7943 & 6.70 & 4.15 & 1.82(15) & 1.57 & 1183 & 1.16 \\
12 & Cl$1227.9$--$1138$ & 0.6357 & 3.76 & 3.76 & 3.29(14) & 1.44 &  648 & 1.13 \\
13 & Cl$1227.9$--$1138$a& 0.5826 & 2.21 & 3.61 & 6.93(13) & 1.40 &  382 & 1.12 \\
14 & Cl$1232.5$--$1250$ & 0.5414 & 6.95 & 3.49 & 2.21(15) & 1.37 & 1199 & 1.11 \\
15 & Cl$1301.7$--$1139$ & 0.4828 & 4.34 & 3.31 & 5.69(14) & 1.33 &  755 & 1.10 \\
16 & Cl$1301.7$--$1139$a& 0.3969 & 2.37 & 3.02 & 1.05(14) & 1.26 &  423 & 1.08 \\
17 & Cl$1353.0$--$1137$ & 0.5882 & 4.33 & 3.63 & 5.16(14) & 1.41 &  746 & 1.12 \\
18 & Cl$1354.2$--$1230$ & 0.7620 & 4.27 & 4.08 & 4.70(14) & 1.54 &  749 & 1.16 \\
19 & Cl$1354.2$--$1230$a& 0.5952 & 2.82 & 3.65 & 1.42(14) & 1.41 &  486 & 1.12 \\
20 & Cl$1411.1$--$1148$ & 0.5195 & 4.54 & 3.43 & 6.27(14) & 1.35 &  785 & 1.11 \\
21 & Cl$1420.3$--$1236$ & 0.4962 & 1.38 & 3.35 & 1.82(13) & 1.34 &  240 & 1.10 \\
\enddata
\tablecomments{(1) cluster field; (2) cluster name; (3) cluster redshift;
  (4,5) infall radius in units of Mpc and units of observed-epoch virial radii;
  (6,7) virial mass evolved to $z=0$ in units of \msun\ and units of observed-epoch
  virial masses; (8,9) velocity dispersion evolved to $z=0$ in units of km~s$^{-1}$
  and units of observed-epoch $\sigma$.}
\end{deluxetable}

\subsection{Number of Infalling Galaxies}\label{sec:n_infall}

We estimate the richness of the infall regions ($R_{\rm 200}<R<R_{\rm infall}$),
which we define as the number of cluster galaxies in the infall region ($N_{\rm
  infall}$) above an absolute $B$-band magnitude of $M_B=-18.9$
(corresponding to $R_{\rm AUTO}\approx 22.9$ for our highest-redshift
cluster at \hbox{$z=0.79$}).

Because we do not have redshifts for every galaxy above this magnitude limit, to estimate $N_{\rm
  infall}$, we use the number of photometric sources in the radial range $R_{\rm
  200}<R<R_{\rm infall}$, $N_{\rm phot}$, multiplied by an estimate
of what fraction of $N_{\rm phot}$ lies within $\Delta z \pm 0.02$ of the
cluster. We determine this fraction from the ratio of
LDP-selected cluster members, $N_{\rm memb}$, to the number of LDP slits,
$N_{\rm slits}$, limiting both to the infall region. This procedure
accounts for the incomplete spatial sampling due to chip gaps and masked
bright stars. We estimate the contamination from field galaxies ($f_{\rm contam}$)
by using the fraction of field galaxies at that cluster redshift; these galaxies
lie at $z_{\rm clus}$ but are observed in fields {\it other} than that particular
cluster (excluding any that lie within another EDisCS cluster; see \S\ref{sec:cmd}).
Therefore, $N_{\rm infall}$ is calculated as
\begin{equation}
N_{\rm infall}=N_{\rm phot} \frac{N_{\rm memb}}{N_{\rm slits}} (1-f_{\rm contam}).
\end{equation}

We also estimate the number of cluster galaxies within $R_{\rm 200}$ at $z=z_{\rm clus}$,
$N_{\rm cluster}$, using the same methodology and magnitude limits, but apply an additional
correction to account for close pairs that the LDP might miss (see Section~\ref{radcomplete}).
To make this correction, we can look at how many of the core cluster
galaxies targeted with VLT/FORS2 were also targeted with the LDP and divide by that
fraction, which is $18\%$.
This correction is slightly underestimated because the slit geometry resulted in a slight undersampling of close pairs within the EDisCS clusters.
However, after applying this correction, our derived values of $N_{\rm cluster}$ agree within the uncertainties to previous estimates made using spectroscopically confirmed EDisCS
galaxies \citep{Poggianti10}.

\begin{figure}
\epsscale{0.8}
\plotone{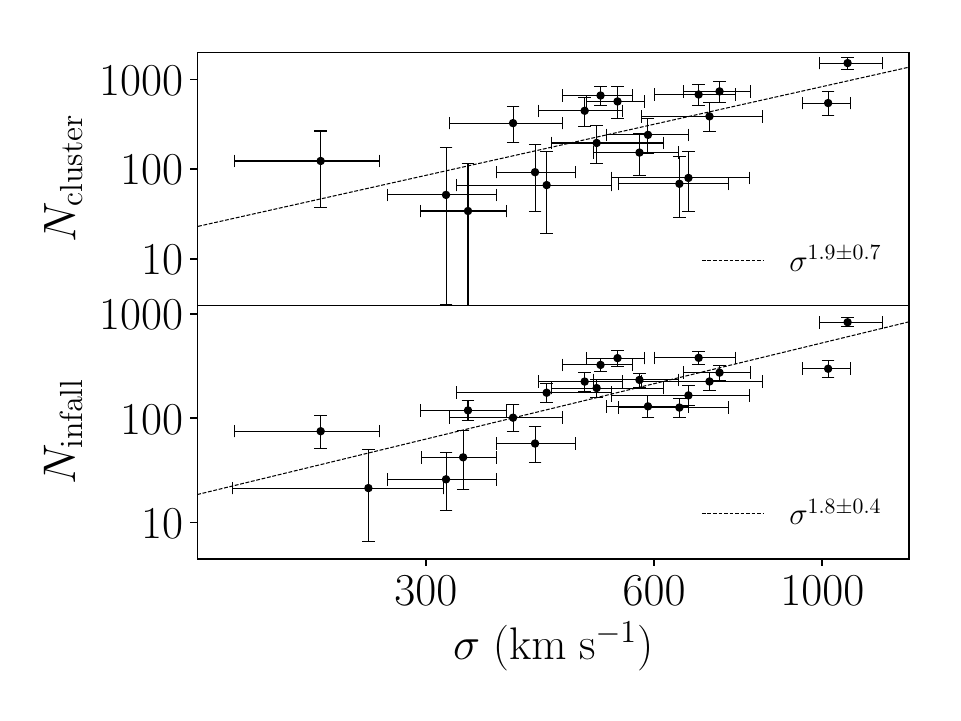}
%\figurenum{8}
\caption{\label{ngal}({\it Top panel}) Number of galaxies within the
  virial radius ($N_{\rm cluster}$) as a function of cluster velocity
  dispersion ($\sigma$). The dashed line is the best fit to the
  data. ({\it Bottom panel}) Same as the top panel, only for galaxies in
  the infall region ($N_{\rm infall}$).}
\end{figure}

In Figure~\ref{ngal}, we present a plot of $N_{\rm cluster}$ and
$N_{\rm infall}$ versus $\sigma$. Errors on $N_{\rm cluster}$ and
$N_{\rm infall}$ are Poissonian \citep[calculated using the equations of][]{Gehrels86},
and errors in $\sigma$ come from \cite{Halliday04} and \cite{MilvangJensen08}. We
perform a linear regression that follows a Bayesian approach and accounts for errors in
both $\sigma$ and $N$, using the IDL routine {\sc linmix\_err.pro} written by
\citet{Kelly07}. We find $N_{\rm cluster}\propto\sigma^{1.9\pm0.7}$ and $N_{\rm
  infall}\propto\sigma^{1.8\pm0.4}$. While the trend for $N_{\rm cluster}$ has a steeper
dependence on $\sigma$ than that of the $z\sim0$ clusters observed by \cite{Finn08}, who
found $N_{\rm cluster}\propto\sigma^{1.4}$, the results are within the uncertainties.
Given the large errors, we find no statistically significant difference in the intercepts
in either region. While more precise measurements are needed to reduce the uncertainty,
our best-fit scalings between $N_{\rm cluster}$ and $N_{\rm infall}$ with $\sigma$ are
comparable.

\begin{figure}
\epsscale{0.8}
\plotone{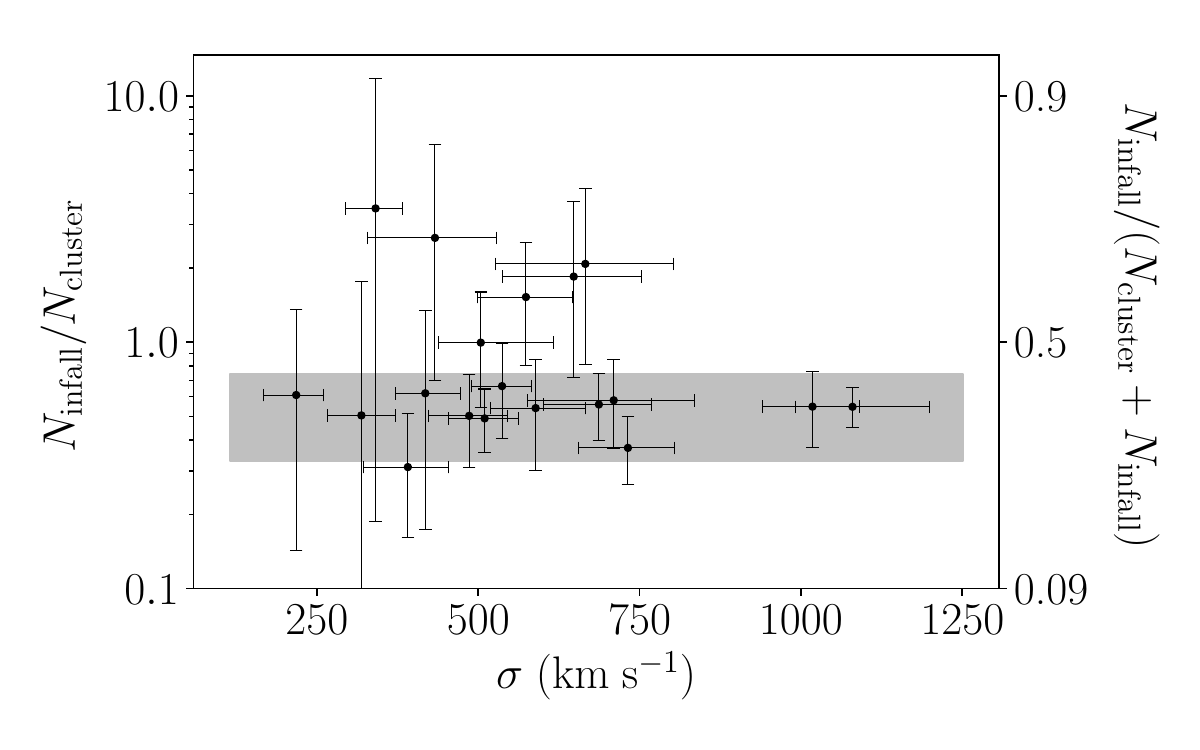}
%\figurenum{9}
\caption{\label{accrete} Ratio of infalling galaxies to cluster
  galaxies as a function of velocity dispersion ($\sigma$). The horizontal
  gray bar comes from converting the expected mass increase based on
  the secondary infall models (Table~5) to galaxy number using $\sigma_8$ (see text).
  No trend is present, such that clusters over this range of $\sigma$ accrete
  proportionally similar numbers of galaxies as they evolve to $z=0$ (although
  they typically range from $\sim0.3$ to $2.0$).}
\end{figure}

Given the similar slopes, we now examine whether clusters of greater mass
accrete proportionally more or fewer galaxies over this redshift interval. Such
behavior would have ramifications for the $\sigma$-dependent increase in S0s as a
fraction of cluster galaxies \citep{Just10}. For example, if more massive
systems accrete a larger percentage of their galaxies at late times
relative to less massive systems, then it could be that the (proportionally
larger) infalling population diluted any increase in the S0 fraction in these
systems, rather than that the less massive systems are intrinsically more
efficient at converting spirals to S0s. In Figure~\ref{accrete}, we compare the
ratio of $N_{\rm infall}$ to $N_{\rm cluster}$ as a function of $\sigma$.
We find that the relative size
of the infalling population does not scale with cluster velocity dispersion.
However, there is a considerable range in the ratio, from $\approx15\%$--$300\%$,
with typical values between $\sim\%30$--$200\%$. 
To determine whether the scatter of $N_{\rm infall}$ or $N_{\rm cluster}$ dominate the scatter in the ratio, we set either the scatter in $N_{\rm infall}$ or $N_{\rm cluster}$ to zero and evaluated the scatter in the ratio. We found that setting the scatter in $N_{\rm infall}$ to zero had minimal effect on the scatter in the ratio and conclude that it is the scatter in $N_{\rm cluster}$ that dominates. This result highlights the importance of accounting for the large variation in cluster properties.

%%
%% Figure 8
%%

%%
%% Figure 9
%%

We now use $N_{\rm cluster}$ and $N_{\rm infall}$ to predict the mass evolution of our
clusters based on the LDP data. We model the correspondence between our clusters
at their observed epoch and at $z=0$ using the model described in \citet{Poggianti06}. In this model,
we account for the enhanced clustering of
galaxies relative to the underlying mass distribution, parameterized by
$\sigma_8$, the RMS fluctuation of galaxies in an $8 h^{-1}$~Mpc sphere relative to
fluctuation in mass, and adopt a recent value, $\sigma_8=0.81$ \citep{Jarosik11}.

From Table~5, we find that the typical increase in mass predicted by our adopted
secondary infall models is $26\%$--$57\%$, or in terms of galaxy number,
$33\%$--$74\%$. This is shown as a gray band in Figure~\ref{accrete}, where it is
consistent with our measured values of $N_{\rm infall}/N_{\rm cluster}$, which are
typically $\sim30$--$150\%$ ($24$--$110\%$ in mass). These values are lower than the
factor of two mass increase predicted for $0.1<z<0.3$ clusters of \cite{Rines13},
although their use of the caustic technique means that their prediction should be higher
than ours, because theirs is for the final cluster mass in the distant future, not
$z=0$. Similarly, \cite{Dressler13} find that among very rich clusters at $z\sim0.4$,
the number of galaxies in infalling groups will roughly double the mass of the
clusters by the present, which is larger than our estimate but still consistent
within the scatter. That we find a comparable mass increase among some of our
clusters that have lower velocity dispersions than their sample
($\approx 800$--$1100$~km~s$^{-1}$) further supports our conclusion from
Figure~\ref{accrete} that more massive systems do not accrete proportionally more
galaxies as they evolve.

Our cluster sample spans redshifts between $z=0.39$ and $z=0.79$ corresponding to galaxy infall times of 4.3 to 6.8 Gyr.
We have investigated the impact that this range of infall times has on the ratio $N_{\rm infall}/N_{\rm cluster}$.
We fit a power law in redshift to the ratio and find a best-fit index of 1.34.
After subtracting off the redshift dependence, we find no significant difference in the mean or the scatter of the cluster ratios.
We conclude that the even with our wide range of infall times, our results are consistent with the secondary infall model.

\subsection{Optical Properties of the Cluster and Infalling Galaxies}\label{sec:cmd}

To estimate the quiescent fraction of galaxies in different environments, we use
the fraction of optically red galaxies. We examine the CMDs of our clusters and
compare the red fractions of core, infalling, and field
galaxies. To construct field samples for each cluster, we select galaxies at the
same redshift but observed in fields {\it other} than the cluster's (excluding any
that overlap in redshift with the EDisCS cluster of that particular field). Note
that (1) this means that there are significantly more galaxies in a given field sample than
in the corresponding cluster, because they are drawn from multiple fields, and (2) we
combine a subset of these field samples in some of the analyses below, so we distinguish
between ``individual field samples" and a ``combined field sample," the reason and
details for which are described below.

%%
%% Figure 10
%%
\begin{figure}
\plotone{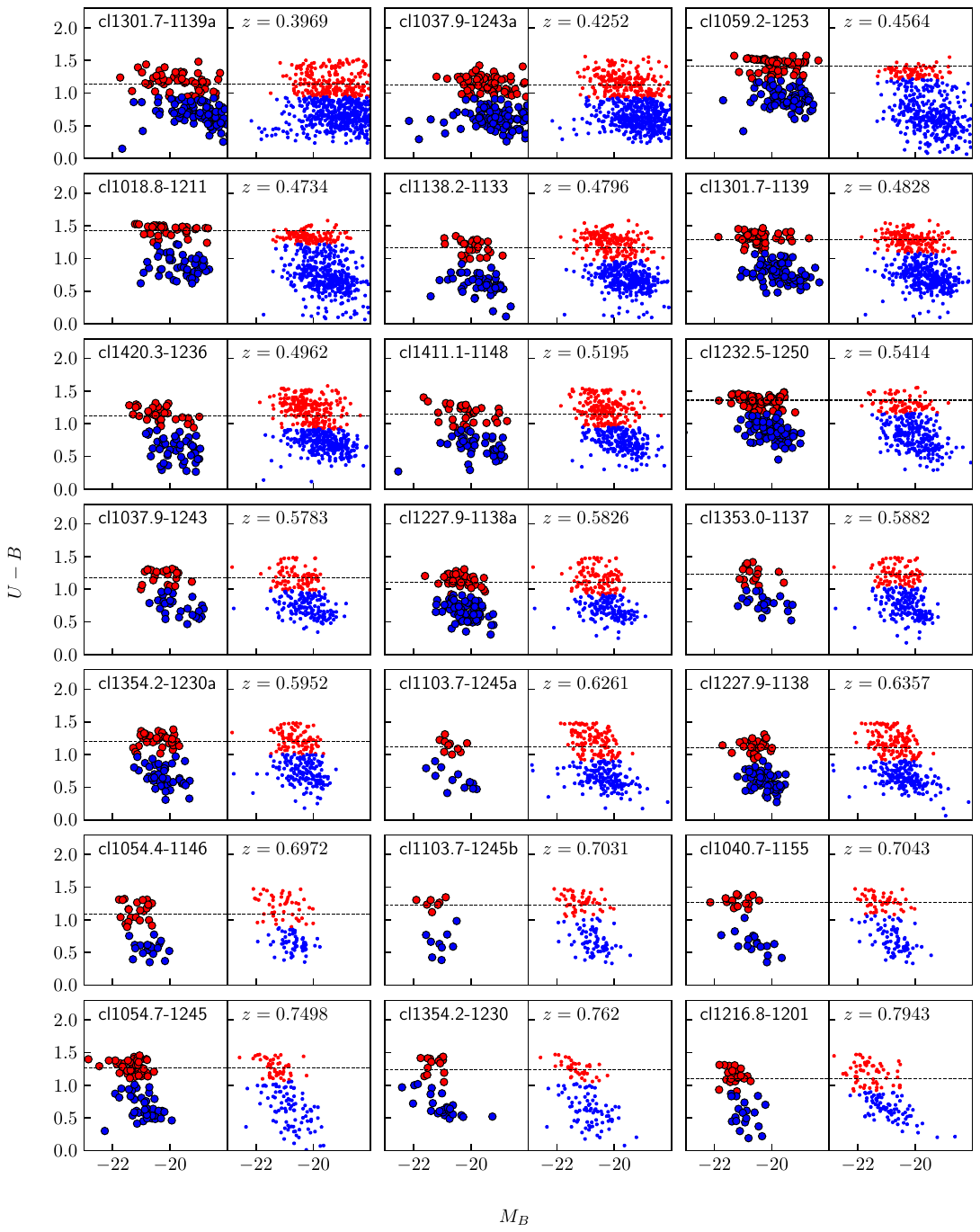}
%\figurenum{10}
\caption{\label{cmd} CMDs for our clusters
  in rest-frame $U-B$ vs. absolute $B$-band magnitude. The left frames show
  LDP-selected cluster members as circles that are colored based on our
  red/blue definition, while the right frames show the individual field
  samples. Dotted lines mark the cluster CMRs. Note that
  because of the way field samples are constructed, they contain many more galaxies than
  their corresponding cluster.}
\end{figure}

For our CMDs, we use rest-frame $U-B$ colors and absolute $B$-band magnitudes calculated
using EAZY \citep{Brammer08}. In
Figure~\ref{cmd} we present observed-frame CMDs for the 21 clusters. 
We measure the color-magnitude relations (CMRs) for all of our clusters by assuming zero 
slope and fitting the WFI $U-B$ colors of the subset of cluster galaxies that have 
FORS2 spectra showing no [OII] emission.
The CMRs measured this way are in agreement with the apparent red sequences of LDP-selected
cluster galaxies (Figure~\ref{cmd}). In what follows, we define galaxies with colors within
$0.2$ of the CMR or redder as {\it red}, while the remaining galaxies are classified as
{\it blue}.

\begin{figure}
\epsscale{1.0}
\plotone{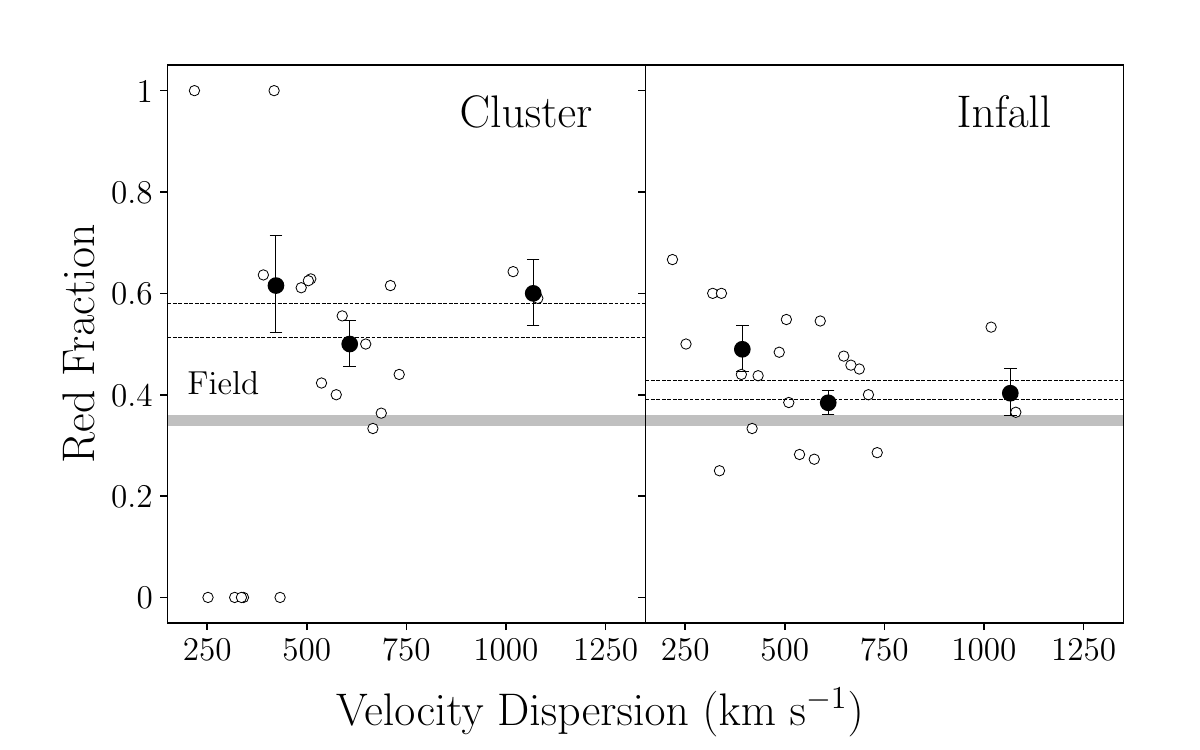}
%\figurenum{12}
\caption{\label{fredvsig} Red fraction as a function of cluster velocity dispersion
  for the cluster and infalling populations. Individual clusters are shown as open circles
  (with error bars suppressed for clarity), while large filled circles are
  the red fractions using all of the galaxies that lie in clusters with 
  $\sigma<500$~km~s$^{-1}$, $500<\sigma<1000$~km~s$^{-1}$, and
  $\sigma>1000$~km~s$^{-1}$. Horizontal dotted lines show the mean $\pm$ the RMS of the 
  overall red fraction for all clusters. Similarly, the overall red fraction of the 
  combined field sample is shown as a shaded horizontal line, with a width corresponding 
  to $\pm$ the RMS of the field red fraction.
  There are no statistically significant trends of red fraction with $\sigma$ in
  either environment. However, the red fraction decreases as one moves from the cores of
  clusters to the field.}
\end{figure}

Because some studies have shown that environmentally 
driven galaxy evolution is correlated with the velocity dispersion
of the group/cluster \citep[e.g.,][]{Poggianti09,Just10}, in Figure~\ref{fredvsig} we
plot the red fraction of the cluster and infalling samples for each cluster as a
function of $\sigma$. To compare to the field, we create a ``combined field sample."
The ``individual field samples" in Figure~\ref{cmd} have galaxies common to more than
one sample. To avoid this multiple counting in the ``combined field sample," we select the
``individual field samples" of eight clusters that span the full redshift range from $0.4$ to
$0.8$ {\it but do not overlap in redshift}. The red fraction for the combined field sample is
also shown in Figure~\ref{fredvsig}. We do not find a correlation between red fraction and
$\sigma$ in the cluster environment, consistent with the findings of \cite{Valentinuzzi11} and \cite{Blanton09}, and
find no correlation in the infalling sample, either. However, the red fraction increases as one
moves from the most isolated environment to the cores of clusters, from $36\pm1\%$ in the field,
to $38\pm2\%$ in the infall region and $55\pm3\%$ in the virial region.

%%
%% Figure 12
%%

The red fraction in the infall regions is slightly elevated relative to the field for the lowest-mass clusters but overall, the sample is consistent with the field.
Within the virial radius, the red fraction is elevated to
$4.3\sigma$ above the infall region and $5.5\sigma$ above the field. Because galaxies
move $\sim$~Mpc distances over $\sim$~Gyr timescales, the quenching of star formation
could begin to occur in the infall region \citep[e.g.,][]{Balogh00}, or even
primarily occur there, with the higher red fraction within $R_{\rm 200}$ owing to the lag
between the start of quenching and the time for its effects to become apparent. With
high-resolution imaging, one would be able to assess whether the infalling red galaxies
exhibit early-type morphologies or perhaps a transitory phase as passive disks.

\cite{Rudnick09} find that the total light on the red sequence for $16$ of the EDisCS clusters
must increase by a factor of $\sim1$--$3$ by $z=0$. We predict that the clusters in our sample
will grow by a factor of $\sim2.1$ in number of galaxies (Figure~\ref{accrete}). Given that the
red fractions within the cluster and infall regions are $55\%$ and $38\%$, respectively, passive
galaxies already identified as such in the infall regions will increase the $z=0$ total red
sequence light by a factor of $\sim1.8$. We conclude
that significant further quenching of blue galaxies in the infall regions as the clusters 
evolve to $z=0$ is not required by our sample. 
Although the uncertainties remain large given the limitations of the current sample and we cannot exclude significant further quenching, this line of reasoning holds promise as a consistency check on quenching models.

``Preprocessing'' has been suggested as a way of transforming galaxies in
locally overdense clumps prior to their incorporation into the cluster
\citep[e.g.,][]{Zabludoff98,Moran07,Kautsch08,Dressler13,Haines13,Lopes13,Cybulski14}.
The elevated red fraction in the infall regions for low-mass clusters, where locally 
overdense clumps are expected to exist, is consistent with preprocessing. In the higher-mass clusters, a mixture of overdense quenching and underdense blue regions may result
in an average red fraction that is barely elevated with respect to the field.
However, this result does not rule out
an additional global mechanism for quenching star formation, one that affects all galaxies at
a given clustercentric radius equally. To explore this scenario further, we measure the amount
of clustering among the galaxies in the infall regions, which will provide more direct
evidence for the association of preprocessing with local overdensities. We measure the
fraction of infalling red/blue galaxies with at least one infalling neighbor of similar
color within a projected distance $d_\theta$, which we denote $F(< d_\theta)$, for values of
$d_\theta$ ranging from $0.15$ to $0.35$~Mpc. A control sample is constructed where we
select galaxies that (1) lie within the infall region of the cluster that they were imaged in; 
(2) are not a member of the cluster that they are imaged in, 
$z^{\mathrm{imaged}}_{cl} - z_{LDP}  > 0.02$; and (3) have a redshift consistent with cluster
member for another cluster in the sample, $z_{\rm cl} - z_{\rm LDP}  < 0.02$.

\begin{figure}
%\epsscale{0.8}
\plotone{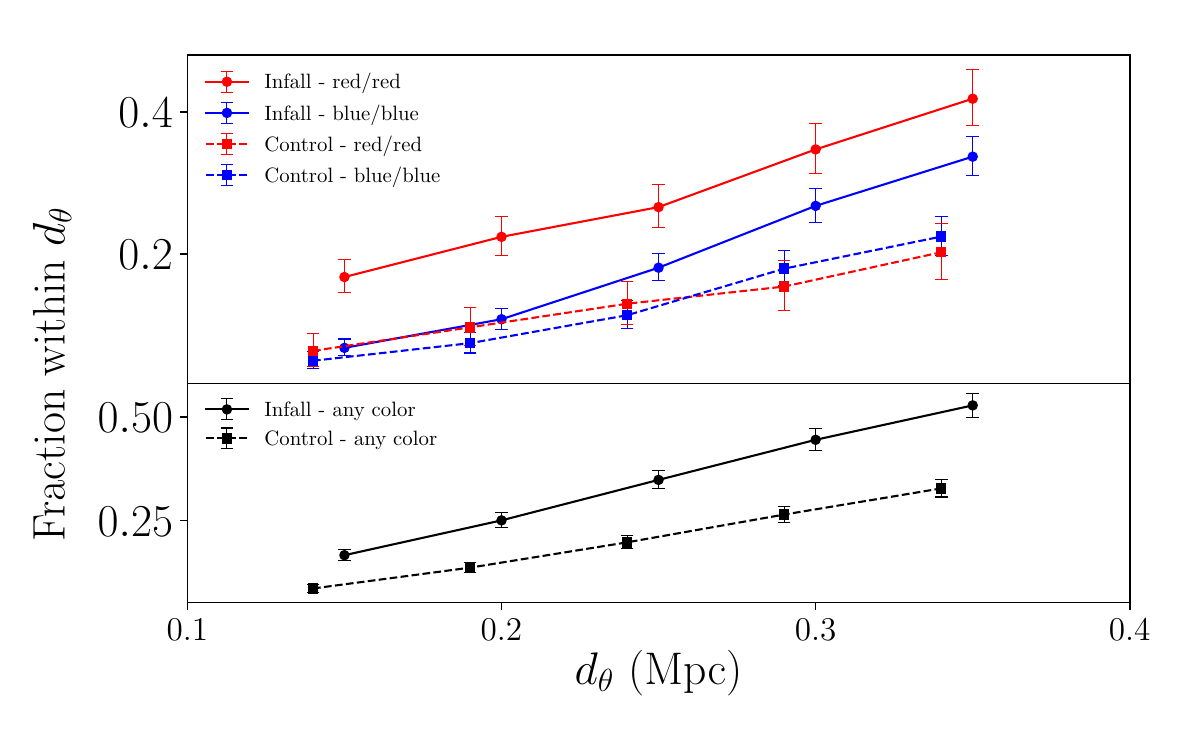}
%\figurenum{13}
\caption{\label{excess}({\it Top panel}) The fraction of infalling red and blue galaxies
  (red and blue circles, respectively) with at least one other infalling red/blue galaxy
  within a projected distance $d_\theta$. Also shown are the fractions for a control
  sample consisting of galaxies at the same redshifts and radial distances as the infall
  regions (red and blue squares). The control data are slightly offset in $d_\theta$ for
  clarity only. ({\it Bottom panel}) Similar to the top panel, only showing the fraction of
  infalling and control galaxies having a neighbor of either color within a distance $d_\theta$.}
\end{figure}

\begin{figure}
%\epsscale{0.8}
\plotone{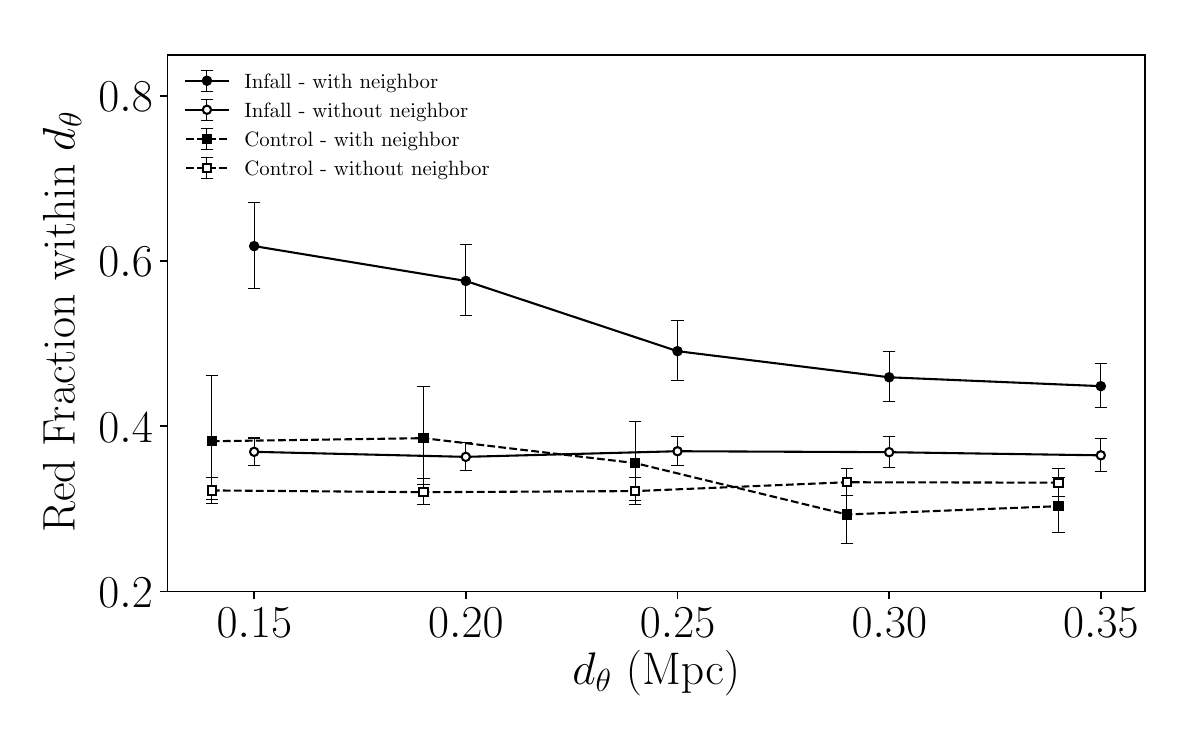}
%\figurenum{14}
\caption{\label{redvneighbors} Red fraction of galaxies with or without a neighbor
  within a projected distance $d_\theta$, for both the infalling and control samples.
  Galaxies with close neighbors are more likely to be red in both samples, but in the
  infall region the enhancement in red fraction is higher than for the control sample.}
\end{figure}

We present these distributions in Figure~\ref{excess}. The clearest result is the elevated fractions of red galaxies that have red neighbors within the infall region relative to either the control or the blue infall galaxies. A second notable finding is that the infalling blue galaxies are not significantly more clustered at small separations ($d_\theta < 0.2$ Mpc) than the control blue galaxies. Alternatively, we measure the red fraction among close pairs of galaxies and 
compare to those without a neighbor (Figure~\ref{redvneighbors}).
From both figures, we find that clustered galaxies are significantly more likely to be red than those without 
a neighbor, and this effect is more significant in the infall regions than 
in the control sample. Evidently, at these length scales the infall regions 
show signs of enhanced clustering of red galaxies, consistent with ``preprocessing", 
in which local overdensities, rather than global environment, quench star formation prior to their incorporation 
into the cluster.

We see the enhanced clustering and elevated red fraction among the infalling sample
(Figures~\ref{excess} and \ref{redvneighbors}) relative to the control (i.e. field)
sample. But despite our definition of clustering, based on having a neighbor within
$d_\theta$, being the same for both samples, the red fractions are different.
Therefore, either the infalling galaxies lie in higher local overdensities, or they
experience an additional effect unrelated to local density. The spatial sampling rate of
our LDP spectra means that we cannot directly compare the numbers of
spectroscopically confirmed galaxies within $d_\theta$ between the two samples. Instead,
we compare (1) the number of photometric sources and (2) the total $R$-band luminosity
within $d_\theta$ to estimate the overdensities. We find that, relative to the control
sample, the infalling cluster members have $\sim3\pm2$ galaxies per Mpc$^2$ more
neighbors and contain $\sim45\%\pm7\%$ more $R$-band luminosity within $d_\theta$. While a
more accurate measurement of local density using a higher sampling of redshifts would be
preferable, this result is consistent with the idea that the infalling galaxies lie in
higher local overdensities than the control galaxies, resulting in more clustering and a
higher red fraction in the infalling sample (Figures~\ref{excess} and \ref{redvneighbors}).

The outlier rate for $Q=4$ galaxies is $12\%$ (\S~\ref{sec:accuracy}), which may
have an impact on the results presented in this section as a result of contamination of
the cluster samples by field galaxies. Given that we found no redshift or color dependence
of the outlier rate, we expect both galaxies at the cluster redshift and field galaxies to
be affected equally by redshift inaccuracy. We use the CMDs to estimate the relative
numbers of cluster and field galaxies and find that the field galaxies constitute
$\approx30\%$ of the total at a given redshift. Therefore, we expect a contamination of
$12\%\times30\%=3.6\%$ field galaxies in the analyses above and do not expect this to be
significant enough to alter our main results.

\cite{Dressler13} found that quiescent and post-starburst (PSB) galaxies are
preferentially found in denser environments, including infalling groups in the
outskirts of rich clusters at $z\sim0.4$. Furthermore, they identified a positive
correlation between the fraction of quiescent and PSB galaxies in the infalling
groups with increasing group mass, which they interpreted as evidence for
``preprocessing." While we are unable to identify PSB galaxies with the LDP
resolution, a prediction based on these results is that our infall regions contain a
higher number of these galaxies than the field. If some fraction of our blue galaxies
are PSB galaxies, then the quiescent plus PSB fraction in the infall regions would be
even more different than the field value. Higher-resolution spectroscopy is needed to
test this hypothesis.

%%
%% Figure 13
%%

%%
%% Figure 14
%%

In addition to environment, star formation is correlated with galaxy stellar mass
\citep[e.g.][]{Kauffmann03,Pasquali09,Peng10}. \cite{Thomas10} find that stellar mass may
even be entirely responsible for the formation of early-type galaxies, at least at
masses $\gtsim 10^{11}$\msun. 
We explore whether the clustered and isolated galaxies (defined by a
separation $d_\theta$) have different mass distributions. Following the prescription of
\citet{Bell03}, we estimate stellar masses using $B$-band mass-to-light ratios, $(M_*/L)_B$,
that are derived from $B-V$ rest-frame colors via
\begin{equation}
\log(M_*/L)_B=1.737(B-V)-0.942,
\end{equation}
after converting our magnitudes to the Vega system. This assumes a diet Salpeter initial
mass function as defined in \cite{Bell01}. Using $M_B=5.45$ for the Sun, a galaxy with
$M_B=-19.5$ and $B-V=1$ has a stellar mass of $\log(M_*/M_\odot)=10.8$. Since our infalling
galaxies span a mass range $\log(M_*/M_\odot)=9$--$12$, we repeat the analyses of
Figures~\ref{excess} and \ref{redvneighbors} but restricting to a narrower mass range,
$\log(M_*/M_\odot)=10$--$11$, which is roughly symmetric about the median mass
($\log(M_*/M_\odot)\approx10.5$). For this narrower mass range, infalling red galaxies are
significantly more likely to have a neighbor than blue galaxies, by $\approx\%12\pm5\%$ compared
to $\approx2\%\pm2\%$ for the control sample. However, we cannot statistically conclude that infalling
galaxies of either color are more likely to have a neighbor of the same color than the control sample,
finding $\Delta F(<d_\theta)\approx2\%\pm2\%$. As when considering the full mass range, the
infalling galaxies with a neighbor have a higher red fraction than isolated galaxies, by
$\approx 50\%\pm12\%$ compared to $\approx 10\%\pm4\%$ for the control sample. An even narrower choice of
masses than this leaves the results qualitatively unchanged, although the number of galaxies becomes
too few to reach statistically significant conclusions like those listed above. While larger numbers of
galaxies would help conclusively rule out a significant mass effect, based on these results we
conclude that the enhanced red fraction among clustered galaxies is consistent with a primarily
preprocessed origin.

Until now we have assumed that all galaxies in the infall regions are falling in for the
first time. However, at these clustercentric radii there exists a population of galaxies that
have already passed through the virial region in the past, so-called ``backsplash" galaxies
\citep{Gill05}. These galaxies may have been quenched on their initial passage (or passages)
through the main body of the cluster, independent of any preprocessing, and therefore must be
accounted for. \cite{Balogh00} suggest that as many as $54\%\pm20\%$ of galaxies at distances between
$R_{\rm 200}$ and $2 R_{\rm 200}$ are members of this backsplash population. Other studies
involving backsplash galaxies focus on distances between the virial radius and $2.5$ times
the virial radius \citep[e.g.,][]{Mamon04,Oman13}.  $N$-body simulations show that satellite galaxies
ejected from the host halo could constitute $\sim 10\%$ of galaxies at $2 R_{\rm 200}$--$5 R_{\rm 200}$
\citep{Wetzel13}. While these studies show that a sizable fraction of backsplash galaxies could
be present in the infall regions, 
these are not expected to be as clustered as 
first-infall galaxies because we expect that their cluster crossing separated galaxies 
that fell in together. We base that expectation on the consideration that infalling groups
are of roughly the same physical scale as the cluster core but of lower mass, and 
therefore we expect tidal effects to dissolve the group. The efficiency and ubiquity of this 
process need to be evaluated quantitatively with simulations. If our conjecture is correct, then
a dominant contribution by backsplash galaxies is
at odds with our findings in Figure~\ref{excess} and \ref{redvneighbors}, in which the red
galaxies are significantly more clustered and the red fraction of clustered galaxies is
significantly higher than those without close neighbors. However, we cannot quantify
the significance of the backsplash population in driving the enhanced quenching we
interpret in the infall regions. It will be quite difficult to disentangle these two
populations.

Matches may also include some associations that are not physical but are rather chance projections. These could include both matches with unassociated galaxies that are cluster members and matches with galaxies beyond the cluster environment. Given our poor redshift resolution, there is no way to identify such cases using our data. Ultimately, a comprehensive analysis of simulated data in a manner that is consistent with our observing methodology should be carried out but is beyond the scope of this paper. The excess found relative to the control in the correlation of red galaxies with other red galaxies suggests that the bulk of the signal is real, given that at these radii the cluster environment is not dominated by red galaxies, but quantitative conclusions will await the full simulations. Given the large uncertainties in our understanding of how well infalling structures survive, any detailed analysis of the data may be 
premature.

%%%%%%%%%%%%%%%%%%%%%%%%%%%%%%%%%%%

\section{Conclusion}\label{sec:conclusion}

We present a spectroscopic survey of $21$ EDisCS clusters at $0.4<z<0.8$
using LDP/IMACS low-resolution spectroscopy. This survey contains 35,280
galaxies (with $1763$ within $\pm0.02$ of the corresponding cluster redshift) and has an
accuracy of $\sigma_z=0.007$.

We have isolated the galaxies in the infall regions of these clusters using the
LDP data and a simple model of secondary infall. The projected distance that
encompasses the infalling galaxy population, $R_{\rm infall}$, agrees to
simulations within $\sim30\%$. The predicted cluster velocity dispersions at
$z=0$ agree with the models of \cite{Poggianti06} to $10\%$. 

With the LDP data, we identified the number of galaxies in the infall
regions and estimate that $\sim30$--$70\%$ of the $z=0$ cluster population
lies outside the virial radius at $z\sim0.6$, a result that is not
sensitive to the mass of the cluster over the range of cluster mass investigated
here. This result demonstrates that studying the infalling population is crucial to
understanding how a significant portion of the galaxy population evolves.
Furthermore, the ratio of the number of infalling galaxies to cluster galaxies is
typically $\sim0.3$--$1.5$. The full range of this ratio is $\approx10$--$300\%$,
highlighting the large cluster-to-cluster variation that exists.

The red fraction in the infall regions is intermediate to that in the field and clusters
for low cluster masses. This suggests that the process of quenching star formation has begun
outside of the virial radius, an effect previously measured at $z\sim0$ \citep{Lewis02,Gomez03}. 
Furthermore, galaxies in the infall regions show enhanced
clustering, and the more highly clustered galaxies also show an elevated red fraction.
These trends are indicative of ``preprocessing," in which galaxy star formation is
shut off in local galaxy overdensities prior to the incorporation of the galaxies
into the cluster, although backsplash galaxies may play a role. Our sample
lies at $z\sim0.6$, before the epoch at which
significant numbers of S0s begin to populate the cores of clusters
\citep{Dressler97,Fasano00}, so it is plausible that the S0s in those cores are the
remnants of quenched infalling galaxies that we see clustered in the infall regions.
Higher-resolution imaging is required to identify the morphologies of these possible
progenitors.

This dataset enables the direct study of galaxies in the infalling regions of
moderate-mass clusters at intermediate redshifts. It clearly demonstrates that further 
studies seeking to understand the mechanisms that halt
star formation in dense environments should target not just the virialized
regions of clusters but the outskirts as well.

\acknowledgements
We thank an anonymous referee for insightful comments that have improved the
content and presentation of this paper.
DWJ and DZ acknowledge financial support from NASA LTSA award NNG05GE82G and GALEX grant
NNX11AI47G. GDL acknowledges financial support from the European Research Council
under European Community's Seventh Framework Programme (FP7/2007-2013)/ERC grant
agreement n. 202781. PJ acknowledges support by the Swiss National Science
Foundation (SNSF).

\appendix
\section{Spatial Maps of Cluster Members}
In Figure~\ref{fig:map1}, we present the spatial map of galaxies in and around our clusters. We mark cluster member galaxies that we classify as red or blue in their respective colors and all other detected galaxies in the field as small black points. We also draw black circles corresponding to the virial radius (inner circle) and the infall radius (outer circle) for each of our clusters

\begin{figure}
\plotone{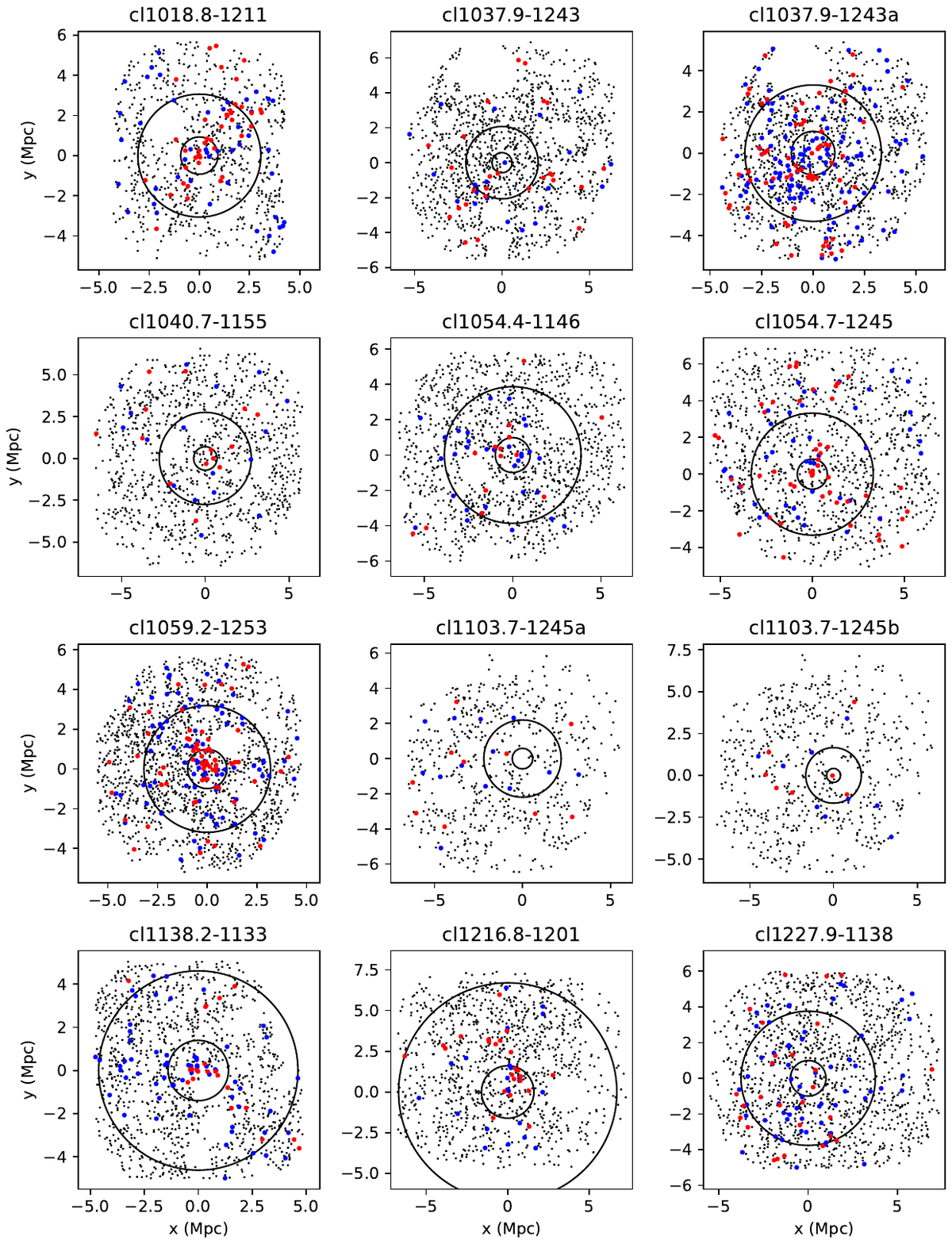}
\caption{\label{fig:map1} Spatial maps of our clusters. Red and blue circles mark cluster members meeting our red/blue definitions. Black circles corresponding to the virial and infall radii are also shown.}
\end{figure}

\begin{figure}
\plotone{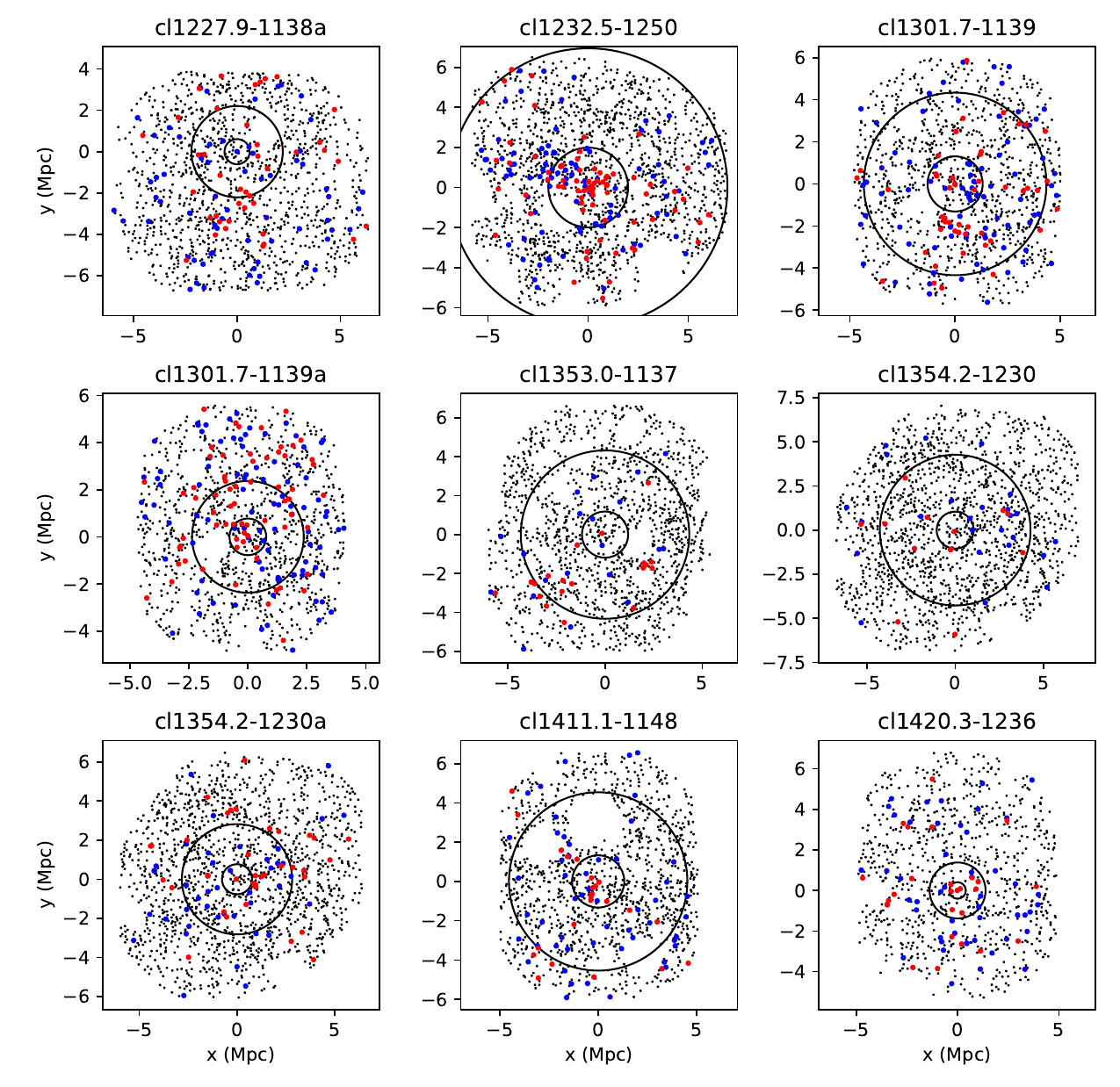}
\figurenum{14}
\caption{Continued}
\end{figure}

%%%%%%%%%%%%%%%%%%%%%%%%%%%%%%%%%%%

\end{document}